\documentclass{aa}
\usepackage{graphicx,color}
\usepackage{bm,natbib,url,wasysym}
\usepackage[varg]{txfonts}
\bibpunct{(}{)}{;}{a}{}{,} 
\graphicspath{{./figs/}{./png/}}

\newcommand{\EQ}{\begin{equation}}
\newcommand{\EN}{\end{equation}}
\newcommand{\EQA}{\begin{eqnarray}}
\newcommand{\ENA}{\end{eqnarray}}

\newcommand{\Eq}[1]{Eq.~(\ref{#1})}

\newcommand{\App}[1]{Appendix~\ref{#1}}
\newcommand{\Sec}[1]{Sect.~\ref{#1}}

\newcommand{\FFig}[1]{\textcolor{red}{Figure~\ref{#1}}}
\newcommand{\Fig}[1]{\textcolor{red}{Fig.~\ref{#1}}}
\newcommand{\Figs}[2]{\textcolor{red}{Figs.~\ref{#1} and \ref{#2}}}
\newcommand{\Figss}[2]{\textcolor{red}{Figs.~\ref{#1}--\ref{#2}}}

\newcommand{\Tab}[1]{\textcolor{magenta}{Table~\ref{#1}}}
\newcommand{\Tabs}[2]{\textcolor{red}{Tables~\ref{#1} and \ref{#2}}}
\newcommand{\Tabss}[2]{\textcolor{red}{Tables~\ref{#1}--\ref{#2}}}
\newcommand{\bra}[1]{\langle #1\rangle}

\newcommand{\mean}[1]{\overline #1}
\newcommand{\meanf}[1]{\overline #1{}^{\rm fil}}
\newcommand{\meanrho}{\overline{\rho}}

{}
{}
\newcommand{\meanFFFF}{\overline{\mbox{\boldmath ${\cal F}$}}{}}{}

\newcommand{\meanSSSS}{\overline{\mbox{\boldmath ${\mathsf S}$}} {}}

\newcommand{\meanSSS}{\overline{\mathsf{S}}}
{}
{}
{}
{}
{}
{}
{}
{}
\newcommand{\meanAA}{\overline{\mbox{\boldmath $A$}}{}}{}
\newcommand{\meanBB}{\overline{\mbox{\boldmath $B$}}{}}{}
{}
{}
{}
{}
{}
{}
{}
{}
\newcommand{\meanJJ}{\overline{\mbox{\boldmath $J$}}{}}{}
{}
\newcommand{\meanUU}{\overline{\mbox{\boldmath $U$}}{}}{}

{}
{}
{}

\newcommand{\meanB}{\overline{B}}

\newcommand{\meanU}{\overline{U}}

{}

{}
{}

\newcommand{\PCn}{{\sc Pencil Code}}
\newcommand{\zzz}{\hat{\mbox{\boldmath $z$}} {}}

\newcommand{\gggg}{\mbox{\boldmath $g$} {}}

\newcommand{\bb}{\bm{b}}
\newcommand{\BB}{\bm{B}}

\newcommand{\uu}{\mbox{\boldmath $u$} {}}
\newcommand{\UU}{\mbox{\boldmath $U$} {}}

\newcommand{\JJ}{\mbox{\boldmath $J$} {}}

\newcommand{\AAA}{\mbox{\boldmath $A$} {}}

\newcommand{\ff}{\mbox{\boldmath $f$} {}}

\newcommand{\FF}{\mbox{\boldmath $F$} {}}

\newcommand{\grav}{\mbox{\boldmath $g$} {}}
\newcommand{\nab}{\mbox{\boldmath $\nabla$} {}}
\newcommand{\OO}{\bm{\Omega}}

\newcommand{\SSSS}{\mbox{\boldmath ${\sf S}$} {}}

\newcommand{\erf}{{\rm erf}}

\newcommand{\meanDD}{{\overline{\rm D}} {}}
\newcommand{\DD}{{\rm D} {}}

\newcommand{\dd}{{\rm d} {}}

\newcommand{\const}{{\rm const}  {}}

\def\degr{\hbox{$^\circ$}}
\def\la{\mathrel{\mathchoice {\vcenter{\offinterlineskip\halign{\hfil
$\displaystyle##$\hfil\cr<\cr\sim\cr}}}
{\vcenter{\offinterlineskip\halign{\hfil$\textstyle##$\hfil\cr<\cr\sim\cr}}}
{\vcenter{\offinterlineskip\halign{\hfil$\scriptstyle##$\hfil\cr<\cr\sim\cr}}}
{\vcenter{\offinterlineskip\halign{\hfil$\scriptscriptstyle##$\hfil\cr<\cr\sim\cr}}}}}

\def\Ta{\mbox{\rm Ta}}

\def\Co{\mbox{\rm Co}}

\def\Pm{\mbox{\rm Pr}_M}
\def\Rm{\mbox{\rm Re}_M}
\def\Tat{\mbox{\rm Ta}_{\rm t}}

\def\Rey{\mbox{\rm Re}}

\def\Co{\mbox{\rm Co}}

\def\cs{c_{\rm s}}

\def\qpz{q_{\rm p0}}
\def\qp{q_{\rm p}}

\def\betap{\beta_{\rm p}}
\def\betamin{\beta_{\min}}
\def\betastar{\beta_{\star}}
\def\Peff{{\cal P}_{\rm eff}}
\def\Peffmin{{\cal P}_{\rm eff}^{\rm min}}
\def\Pmin{{\cal P}_{\rm eff}^{\rm min}}

\def\qs{q_{\rm s}}

\def\qg{q_{\rm g}}

\def\kf{k_{\rm f}}
\def\HP{H_p}

\def\urms{u_{\rm rms}}

\def\nut{\nu_{\rm t}}

\def\nuT{\nu_{\rm T}}

\def\etat{\eta_{\rm t}}
\def\etatz{\eta_{\rm t0}}

\def\etaT{\eta_{\rm T}}

\def\Beq{B_{\rm eq}}
\def\Beqz{B_{\rm eq0}}
\def\Bfm{\overline{B}{}_z^{\rm fil\,max}}
\def\tautd{\tau_{\rm td}}
\def\tauto{\tau_{\rm to}}

\def\csq{c_{\rm s}^2}
\def\Bfm{\overline{B}{}_z^{\rm fil\,max}}

\def\half{{\textstyle{1\over2}}}

\def\onethird{{\textstyle{1\over3}}}

\newcommand{\G}{\,{\rm G}}

\newcommand{\kG}{\,{\rm kG}}

\newcommand{\Mm}{\,{\rm Mm}}

\begin{document}

\titlerunning{Magnetic bipoles in rotating turbulence with coronal envelope}
\authorrunning{Losada et al.}

\title{Magnetic bipoles in rotating turbulence with coronal envelope}

\author{I. R. Losada \inst{1,2,3} \and J. Warnecke\inst{4,1}
\and A. Brandenburg \inst{1,2,5,6}
\and N. Kleeorin \inst{7,1}
\and I. Rogachevskii \inst{7,1}}
\institute{
Nordita, KTH Royal Institute of Technology and Stockholm University,
SE-10691 Stockholm, Sweden\\
\email{illa.rivero.losada@gmail.com}\label{inst1}
\and Department of Astronomy, AlbaNova University Center,
Stockholm University, SE-10691 Stockholm, Sweden \label{inst2}
\and Nordic Optical Telescope, La Palma, Canary Islands, Spain
\and Max-Planck-Institut f\"ur Sonnensystemforschung,
Justus-von-Liebig-Weg 3, D-37077 G\"ottingen, Germany \label{inst3}
\and JILA and Department of Astrophysical and Planetary Sciences,
Box 440, University of Colorado, Boulder, CO 80303, USA \label{inst4}
\and Laboratory for Atmospheric and Space Physics,
Box 590, University of Colorado, Boulder, CO 80303, USA \label{inst5}
\and Department of Mechanical
Engineering, Ben-Gurion University of the Negev, POB 653,
Beer-Sheva 84105, Israel \label{inst6}
}
\date{\today,~ $ $Revision: 1.416 $ $}
\abstract{
The formation mechanism of sunspots and starspots is not fully understood.
It is a major open problem in astrophysics.
}{%
Magnetic flux concentrations can be produced by the negative effective
magnetic pressure instability (NEMPI).
This instability is strongly suppressed by rotation.
However, the presence of an outer coronal envelope was previously
found to strengthen the flux concentrations and make them more prominent.
It also allows for the formation of bipolar regions (BRs).
It is important to understand
whether the presence of an outer coronal envelope also changes
the excitation conditions and the rotational dependence of NEMPI.
}{%
We use direct numerical simulations and mean-field simulations.
We adopt a simple two-layer model of turbulence that mimics the jump
between the convective turbulent and coronal layers below and above the
surface of a star, respectively.
The computational domain is Cartesian and located at a certain latitude
of a rotating sphere.
We investigate the effects of rotation on NEMPI by changing the Coriolis number,
the latitude, the strengths of the imposed magnetic field, and the box
resolution.
}{%
Rotation has a strong impact on the process of BR formation.
Even rather slow rotation is found to suppress their formation.
However, increasing the imposed magnetic field strength also makes the
structures stronger and alleviates the rotational suppression somewhat.
The presence of a coronal layer itself does not significantly reduce
the effects of rotational suppression.
}{}
\keywords{Magnetohydrodynamics (MHD) -- turbulence -- dynamo -- Sun:
  magnetic fields -- Sun: rotation-- Sun: activity
}

\maketitle

\section{Introduction}

The solar dynamo operates in hydromagnetic turbulence in the presence of strong
stratification---especially near the surface.
The stratification can lead to a secondary instability, in
addition to the primary dynamo instability, and can concentrate the field
further into spots.
This instability was discovered by \cite{Kleeorin89,KRR1990}
and applied to explain sunspot formation
and other hydromagnetic processes
in the Sun \citep{KMR93,KMR96,KR94,RK07}.
In the last 10 years, direct numerical simulations (DNS) have demonstrated that this
instability---the negative effective magnetic pressure instability (NEMPI)---is
able to concentrate magnetic fields in different physical environments.

Since its first detection in a stably stratified isothermal setup with
a weak horizontal \citep{BKKMR11} and a weak vertical \citep{BKR2013}
magnetic field, NEMPI demonstrated its operation in polytropic
stratification \citep{LBKR2014} and turbulent convection
\citep{KaBKMR12,KBKKR16}, in the presence of weak
rotation \citep{LBKMR2012,LBKR2013a}, and with a coronal envelope
\citep{WLBKR2013,WLBKR2015}.
In this context, a ``weak'' magnetic field is one that is
of subequipartition strength, but already dynamically important.
Likewise, weak rotation means that the angular velocity is small compared
with the inverse correlation time of the turbulence,
although it is already dynamically important.
The coronal envelope is crucial for producing a bipolar region (BR)
in the domain.
Here we define BRs as structures that are much larger than the
size of individual turbulent eddies.
The emergence of opposite polarities is a consequence
of zero total vertical flux across a horizontal surface.
This implies that regions with weak
large-scale magnetic field are separated by
regions with strong fields
of opposite magnetic polarity.

In principle, spot structures can also be tripolar or quadrupolar
(for example AR~11158), but those cases are rare. Interestingly,
these regions allow one to the determine the magnetic helicity in the
space above by only using the surface magnetic field \citep{BB18}.
This property may help tying the nature of dynamo-generated subsurface
magnetic fields to observations.
A similarly useful tool is helioseismology, which may allow for a detection
of subsurface magnetic fields in the days prior to active regions formation;
see \cite{SRB16} for such work providing evidence for a gradual buildup
of active regions rather than a sudden buoyant emergence from deep down.

The coronal envelope allows the orientation of a weak imposed horizontal
magnetic field to change locally due to the interface between the turbulence zone
and the coronal envelope so that it can attain a vertical component.
On the other hand, rotation plays against the instability (i.e., NEMPI),
which cannot survive above a certain critical rotation rate.
This critical rate was found to be surprisingly
small \citep{LBKMR2012,LBKR2013a}.

It should be mentioned that other types of surface magnetic flux
concentrations have been seen in a number of different circumstances,
all of which share the presence of a strong density stratification.
There is first of all the phenomenon of magnetic flux segregation into
weakly convecting magnetic islands within nearly field-free convecting
regions \citep{TaoWeiss98}.
This has also been seen in several recent high resolution and high
aspect ratio simulations \citep{KapyBeijing12,KBKKR16} and perhaps also
in those of \cite{MS16}.
This process may explain the formation of flux concentrations seen in
the simulation of \cite{SN2012}, in which an unstructured magnetic field
of $1\kG$ is allowed to enter the computational domain at the bottom.
These simulations include realistic surface physics in a domain $96\Mm$
wide and $20\Mm$ deep, so, again, the aspect ratio is large
and there is significant scale separation.
Another approach is the let a flux tube rise from the bottom of the computational
domain to simulate flux emergence at the surface
\citep{FFD93,Fan01,Arc04,Arc05,FAZS17};
see \cite{Fan09} for a review.
Similar results, but with realistic surface physics,
have been obtained by \cite{CRTS10} and \cite{rempel2014}, who let a
semi-torus of magnetic field advect through the lower boundary.
Their simulations showed that the magnetic field is able to rise through
the top $16\Mm$ of the convection zone to form spots; see the reviews by
\cite{CI14} and \cite{SAP14}.
In their simulations, however, flux emergence is significantly faster
than in the real Sun \citep{BSBCGLR16}.
Recently, \cite{CRF17} were able to reproduce a complex sunspot
emergence using a modified magnetic flux bundle from the dynamo
simulation of \cite{Fan2014} in a spherical shell
and inserting it into a setup similar to that of \cite{rempel2014}.

The process of magnetic flux concentration in convection could be related
to the magnetic suppression of the convective heat flux, which, again,
could lead to a large-scale instability \citep{KitchatinovMazur2000}.
On the other hand, \cite{KKWM2010} explain the formation of magnetic flux
concentrations in their radiation-hydromagnetic simulations as being
confined by the random vorticity associated with convective downdrafts.
However, the process seen in their simulations seems to be similar to
flux concentrations found in the downdraft of the axisymmetric simulations
of \cite{GW81}.
There is also a process known as convective collapse \citep{Parker78},
which can lead to a temporary concentration of field from a weaker, less
concentrated state to a more concentrated collapsed state, e.g., from
$1270\G$ to $1650\G$ in the specific calculations of \cite{Spruit79}.
However, the collapsed state is not in thermal equilibrium, so the
system will slowly return to an uncollapsed state.
This effect may be related to the ionization physics, which can strongly
enhance the resulting concentrations \citep{BhatBrandenburg15}.

Strong magnetic flux concentrations have also been seen in simulations
where a large-scale dynamo is responsible for generating magnetic field.
The dynamo arises as a result of helically driven turbulence in the lower
part of the domain, while in the upper part turbulence is nonhelically driven.
In the upper part, the magnetic field displays the formation of strongly
concentrated BRs.
This was seen in DNS in both Cartesian domains
\citep{MBKR14,Jabbari16,Jabbari2017} and spherical shells
\citep{Jabbari15}.

The relation to NEMPI is unclear in some of these cases, because NEMPI
can be excited when the magnetic field is below the equipartition value
of the turbulence.
A negative effective magnetic pressure is possible in somewhat
deeper layers and at intermediate times, and that may be important
in initializing the formation of magnetic flux concentrations.
In particular, it would lead to downward suction along vertical
magnetic field lines which creates an underpressure in the upper parts
and results in an inflow.
The latter causes further concentrations in the upper parts.
This was clearly seen in the axisymmetric mean-field simulations (MFS)
of \cite{BGJKR14}; see \cite{BRK2016} and \cite{LWGRBKR17} for recent
reviews.

In the present work, we consider a setup similar to that of
\cite{WLBKR2013,WLBKR2015}.
There, turbulence of an isothermal gas is
forced in the lower part of a horizontally periodic domain, while the
upper part is left unforced and subject to the response from the
dynamics of the lower part.
This approach has been used to study the effect of a coronal
envelope on the dynamo \citep[e.g.][]{WB2010,WBM2011,WKMB13,WKKB16}
and the formation of coronal ejections \citep{WBM12,WKMB12}.
In these simulations, the simple treatment of the coronal envelope does
not allow for a low plasma $\beta_{\rm c}$ as in the solar corona,
where $\beta_{\rm c}$ is the ratio of magnetic pressure to gas
pressure. However, also in the solar corona the value of
$\beta_{\rm c}$ is not extremely small \citep[e.g.][]{PWCC15} and plasma flows
can play an important role for the formation of loop structures \citep{WCBP17}.

We include the Coriolis force to examine the effects of rotation in
the presence of our simplified corona to study whether this facilitates
the development and detectability of NEMPI, and whether it changes the
critical growth rate above which NEMPI is suppressed.
We also study the dependence on latitude, as well as the dependence on
the numerical resolution.
Finally, we compare our solutions with corresponding MFS, in which
a prescribed effective (mean-field) magnetic pressure operates only
beneath the surface, but not in the coronal layer.

\section{The model}

\subsection{DNS}
\label{Sec:DNS}

We use the same two-layer model as \cite{WLBKR2013,WLBKR2015}.
We considered a Cartesian domain with forced turbulence in the
lower part (referred to as turbulent layer), and a more quiescent
upper part (referred to as coronal envelope).
We further adopt an isothermal equation of state and
solve the equations for the velocity $\UU$,
the magnetic vector potential $\AAA$, and the density $\rho$.
We adopt units for the magnetic field such that
the vacuum permeability is unity.
Here we extend this model by including the presence of rotation with an
angular velocity $\Omega$,
\begin{equation}
{\DD\UU\over\DD t}=-2\OO\times\UU
-\cs^2\nab\ln\rho+\grav+{1\over\rho}\JJ\times\BB+\Theta_w(z)\ff+\FF_\nu,
\label{DUDt}
\end{equation}
\begin{equation}
{\partial\AAA\over\partial t}=\UU\times\BB-\eta\JJ,
\end{equation}
\begin{equation}
{\partial\rho\over\partial t}=-\nab\cdot(\rho\UU),
\end{equation}
where $\DD/\DD t=\partial/\partial t+\UU\cdot\nab$ is the advective
derivative, $\BB=\BB_0+\nab\times\AAA$ is the magnetic field,
$\BB_0=(0,B_0,0)$ is a weak imposed uniform field in the $y$ direction,
$\JJ=\nab\times\BB$ is the current density,
$\FF_\nu=\nab\cdot(2\nu\rho\SSSS)$ is the viscous force,
$\nu$ is the kinematic viscosity,
$\eta$ is the magnetic diffusivity,
$\grav$ is the gravitational acceleration,
${\sf S}_{ij}=\half(\partial_j U_i+\partial_i U_j)-\onethird\delta_{ij}\nab\cdot\UU$
is the traceless rate-of-strain tensor, and $\ff$ is a forcing function
that consists of random, white-in-time, plane, nonpolarized waves within
a certain narrow interval around an average wavenumber $\kf$.
It is modulated by a profile function $\Theta_w(z)$,
\begin{equation}
\label{eq:error}
\Theta_w(z)=\half\left(1-\erf{z\over w}\right),
\end{equation}
that ensures a smooth transition between unity in the lower layer and
zero in the upper layer.
Here $w$ is the width of the transition.
The angular velocity vector $\OO$ is quantified by its modulus
$\Omega$ and colatitude $\theta$, such that
\EQ
\OO=\Omega\left(-\sin\theta, 0, \cos\theta\right).
\EN

Following \cite{WLBKR2013}, the domain used in the DNS is
$L_h\times L_h\times L_z$, where $L_h=2\pi$ and $L_z=3\pi$ with
$-\pi\leq z\leq 2\pi$.
This defines the base horizontal wavenumber $k_1=2\pi/L_h$,
which is set to unity in our model.
Our Cartesian coordinate system $(x,y,z)$ corresponds to a local
representation of a point on a sphere mapped to spherical
coordinates $(r,\theta,\phi)\to(z,x,y)$, where $r$ is radius,
$\theta$ is colatitude, and $\phi$ is longitude.
Similar to earlier work \citep{Kemel12b,Kemel13a}, we use
in all cases $\kf=30\,k_1$ and $\nu=10^{-4}\,\cs/k_1$ with the sound
speed $\cs$.
The normalized gravity is given by $g H_\rho/\csq=1$, which is just
slightly below the value of $1.2$ that was found to maximize the
amplification of magnetic field concentrations \citep{WLBKR2015}.
Here $H_\rho$ is the density scale height.
As in most of our earlier work, we use $k_1 H_\rho=1$, so the
vertical density contrast is $\exp(L_z/H_\rho)=\exp3\pi\approx12,000$.
For the width of the profile functions in the DNS and MFS,
we use $k_1 w=0.05$.

For the rms velocity, we will use the averaged value in the turbulent
layer defined as: $\urms=\bra{\UU^2}_{xy;z\le 0}^{1/2}$.
We normalize the magnetic field by its equipartition value,
$\Beq=\sqrt{\rho} \, \urms$, using either the $z$ dependent value of
$\Beq$ or the value at the surface at $z=0$, i.e., $\Beqz\equiv\Beq(0)$.

Our simulations are characterized by the magnetic and fluid Reynolds numbers,
\EQ
\Rm=\urms/\eta\kf,\quad
\Rey=\urms/\nu\kf,
\EN
respectively, the Coriolis number
\EQ
\Co=2\Omega\tau,
\label{CoriolisNum}
\EN
where $\tau=1/\urms\kf$ is the eddy turnover time,
and the colatitude $\theta$ of our domain is positioned on the sphere.
In the following, we use $\Rm\approx14$ and $\Rey\approx29$,
so the magnetic Prandtl number is $\Pm=\nu/\eta=0.5$.
These values of the magnetic and fluid Reynolds numbers
are based on the forcing wavenumber, which is rather high
($\kf/k_1=30$).
Thus, the values of $\Rm$ and $\Rey$ based on the wavenumber of the
domain would be 420 and 870, respectively, and those based on the size
of the domain, which are larger by another factor of $2\pi$, would be
2640 and 5470, respectively.
The definitions of these Reynolds numbers must therefore be kept
in mind when comparing with other work.
In our definition, the magnetic Reynolds number required for the
effective magnetic pressure to be negative must be larger than a
critical value of about three \citep{BKKMR11,BKKR12,KaBKMR12,WLBKR2015}.
For small $\Rm$, the effective magnetic pressure can only be positive
\citep{RKS12,BKKR12}.
In our work, we choose $\Rm$ to be about ten times supercritical.
With the choice of $\Rey\approx29$ and $\Pm=0.5$, we exclude that the excitation of a
small-scale dynamo influences our results. As shown in
\cite{WLBKR2015}, $\Pm=1$ is needed to excite a small-scale dynamo
in this setup.

In this work, time is often expressed in units of the turbulent diffusive
time, $\tautd=\etatz k_1^2$, where $\etatz=\urms/3\kf$ is an estimate
of the turbulent magnetic diffusivity.
As in earlier work \citep{WLBKR2013,WLBKR2015}, we use the Fourier
filtered magnetic and velocity fields as diagnostics for characterizing
large-scale properties of the solutions.
Our Fourier filtered fields are denoted by an overbar and the superscript
$\rm fil$, i.e., $\meanf{B_z}$ for the vertical magnetic field.
This includes contributions with horizontal wavenumbers below $\kf/2$.
This filtering wavenumber is the same as that used in \cite{WLBKR2013},
but the cutoff wavenumber is three times larger than the filter value
$\kf/6$ used by \cite{BKR2013} and \cite{WLBKR2015}.
This is appropriate here because, owing to the nature of our BRs,
where the spots tend to be close together, they would not be
well captured when the averaging scale is too large or the filtering
wavenumber too small.
The corresponding spectral magnetic energy contained in the
vertical magnetic field, $B_z$, is $E_{\rm M}^z$ and obeys
$\int E_{\rm M}^z\,\dd k=\bra{B_z^2}/2$.
Of particular interest is the energy per logarithmic wavenumber
interval, $2k_\ast E_{\rm M}^z(k_\ast)$, which we usually evaluate at
$k_\ast/k_1=2$, where the energy reaches a maximum.
(The factor $2$ in front of $kE_{\rm M}^z$ compensates for the 1/2 factor
in the definition of the energy.)
For the velocity, however, we find that $\kf/6$ is the appropriate
filtering wavenumber.
Therefore, we filter the velocities on a larger scale than the magnetic
field.

We compute growth rates and magnetic energies as averages over a certain
time interval.
We compute error bars as the largest departure from any one third
of the full time interval used for computing the average.
In some cases, those error estimates were themselves unreliable.
In such exceptional cases we have replaced it by the average error
for other similar simulations.

We use resolutions between $192\times192\times384$ and
$1152\times1152\times2304$ meshpoints in the $x$, $y$,
and $z$ directions, respectively.
We adopt periodic boundary conditions in the $xy$ plane, a stress-free
perfect conductor condition on the bottom boundary, and a stress-free
vertical field condition on the top boundary.

\subsection{MFS}

In this section, we state the relevant equations for the mean-field
description of NEMPI in a system with coronal envelope
in the presence of rotation.
The relevant equations have been obtained by
\cite{Kleeorin89,KRR1990,KMR96} and \cite{KR94}
through ensemble averaging.
In practical applications, these averages should be replaced by spatial
averages, but their precise nature depends on the physical circumstances
and could be planar (e.g., horizontal) or azimuthal (e.g., around a
flux tube), or some kind of spatial smoothing.
In the present case with inclined stratification, rotation vectors, and
mean magnetic field vectors, we expect the formation of bipolar structures
that cannot be described by simple planar or azimuthal averages.
The only meaningful average is a smoothing operation that could preserve
such structures.
In the following, we denote the dependent variables of our MFS by overbars.
For the analysis of our DNS, on the other hand, we use sometimes Fourier
filtering and sometimes horizontal averaging and denote them also by
an overbar.
In those cases, a corresponding comment will be made, and in the particular
case of Fourier filtering, the variable will be additionally denoted by the
superscript ``fil''.

In the MFS, the equations for the mean velocity $\meanUU$, mean vector
potential $\meanAA$, and mean density $\meanrho$, are given by
\begin{equation}
{\meanDD\,\meanUU\over\DD\, t}=-2\OO\times\meanUU
-\cs^2\nab\ln\meanrho+\grav+\meanFFFF_{\rm M}+\meanFFFF_{\rm K},
\label{dUmean}
\end{equation}
\begin{equation}
{\partial\meanAA\over\partial t}=\meanUU\times\meanBB-(\etat+\eta)\meanJJ,
\end{equation}
\begin{equation}
{\partial\meanrho\over\partial t}=-\nab\cdot(\meanrho\meanUU),
\end{equation}
where $\meanDD/\DD t=\partial/\partial t+\meanUU\cdot\nab$
is the advective derivative based on $\meanUU$,
$\etat$ is turbulent magnetic diffusivity,
\EQ
\meanFFFF_{\rm K}=(\nut+\nu)\left(\nabla^2\meanUU+\onethird\nab\nab\cdot\meanUU
+2\meanSSSS\nab\ln\meanrho\right)
\EN
is the total (turbulent plus microscopic) viscous force
with $\nut$ being the turbulent viscosity,
$\meanSSS_{ij}=\half(\meanU_{i,j}+\meanU_{j,i})
-\onethird\delta_{ij}\nab\cdot\meanUU$
is the traceless rate-of-strain tensor of the mean flow
and, as in the DNS, we adopt units for the mean magnetic
field such that the vacuum permeability is unity.
The effective Lorentz force, $\meanFFFF_{\rm M}$,
which takes into account the turbulence contributions,
i.e., the effective magnetic pressure \citep{KMR96,RK07,BRK2016}
and an anisotropic contribution resulting from gravitational
stratification, is given by
\EQ
\meanrho \, \meanFFFF_{\rm M} = \meanJJ\times\meanBB
+\nab\left[\half q_{\rm p}(z,\meanBB)\,\meanBB^2\right]
+\zzz{\partial\over\partial z}\left[q_{\rm g}(z,\meanBB)\,\meanBB^2\right],
\label{efforce}
\EN
where $q_{\rm p}$ and $q_{\rm g}$ are functions that have previously
been determined from DNS \citep{BKKR12}.
\cite{WLBKR2015} found $q_{\rm g}$ to be negative for weak and moderate
stratification (note that the abscissa of their Fig.~6 shows $g/k_1\cs^2$
and not $gH_\rho/\cs^2$, as was incorrectly written).
Thus, there is the possibility of partial cancelation, which we model
here by assuming $q_{\rm p}$ and $q_{\rm g}$ to have the same profile with
\EQ
q_{\rm g}=a_{\rm g} q_{\rm p}\quad\mbox{($a_{\rm g}=\const$)}.
\label{adef}
\EN
\cite{WLBKR2015} determined $q_{\rm p}\beta^2=-q_{\rm
  g}\beta^2=-0.002$, resulting in $a_{\rm g}=-1$ for the same
stratification as in this work ($g/k_1\cs^2=1$).
We note here that the values of $q_{\rm p}$ and $q_{\rm g}$ are
the result of averaging in time and space,
so locally the values can be different, and therefore they do
not need to cancel out locally.
Furthermore, the error estimate by the spread is comparable to the
averaged value, see Fig.~6 of \cite{WLBKR2015}.
We model $q_{\rm p}$ and $q_{\rm g}$ as the product of a part that
depends only on $\beta^2=\meanBB^2/\Beq^2(z)$ and a profile function
$\Theta_w(z)$.
The latter function
varies only along the $z$ direction and it mimics the effects of the coronal
layer, using the same error function as in \Eq{eq:error}, i.e.,
\EQ
q_{\rm p}(\meanBB,z) = q_{\rm p}^{(B)}(\beta^2)\,\Theta_w(z),
\label{parameterization}
\EN
where
\EQ
q_{\rm p}^{(B)}(\beta^2)= {\qpz\over1+\beta^2/\betap^2}
={\beta_\ast^2\over\betap^2+\beta^2}.
\label{param_beta}
\EN
Here $\beta_\ast=\sqrt{\qpz}\,\betap$ is a parameter that can be used
alternatively to $\qpz$ and has the advantage that the growth rate of
NEMPI is predicted to be proportional to it \citep{Kemel13a}.
In the following, we mainly use the parameters found by \cite{LBKR2013a},
namely $\qpz=32$ and $\betap=0.058$, which corresponds to
$\beta_\ast=0.33$.
On one occasion, we also use another parameter combination that will be
motivated by our results presented in \Sec{EffectiveMagneticPressure}
below.
In hindsight, it might have been more physical to use $\Theta_w^2$ in
\Eq{parameterization}, but we know from experience that this would hardly
make a noticeable difference.

We note in passing that, while in both the DNS and the MFS, the coronal
envelope is modeled with the same profile function $\Theta_w(z)$,
in the MFS, it appears underneath the gradient
(inside the effective Lorentz force), while in the DNS it does not.
In the MFS, this leads to an additional term involving the derivative
of $\Theta_w(z)$, which is a gaussian function.
In one of the cases reported below, we checked that the presence of this
term causes a very minor difference in the growth rates.

In addition to adopting the parameterization given by \Eq{parameterization},
the effects of turbulence are modeled in terms of turbulent viscosity $\nut$
and turbulent magnetic diffusivity $\etat$.
Both can be expressed in terms of $\urms$ and $\kf$, whose values are
known from the DNS and give $\nut=\etat\approx10^{-3}\cs H_\rho$.
Thus, $3\etat/\eta=\Rm$ and $3\nut/\nu=\Rey$.
Among the range of other possible mean-field effects, we have included
here only the negative effective magnetic pressure
functions $q_{\rm p}$ and $q_{\rm g}$.

As already noted by \cite{LBKMR2012}, the usual Coriolis number
is not the relevant quantity characterizing the relative importance
of rotation on NEMPI.
A more meaningful quantity is the ratio of $2\Omega/\lambda_{\ast0}$,
where $\lambda_{\ast0}$ is the nominal
value of the growth rate
\EQ
\lambda_{\ast0}=\beta_\ast\urms/H_\rho,
\label{lamast0}
\EN
which was found to be comparable to the growth rate of NEMPI
\citep{LBKMR2012}.
Therefore, the Coriolis number given by \Eq{CoriolisNum},
can be rewritten as
\EQ
\Co={2\Omega\over\urms\kf}
={2\Omega\over\lambda_{\ast0}}\,{\beta_{\ast}\over \kf H_\rho}.
\label{Omlam}
\EN
With $\urms\approx0.1\cs$ and $k_1 H_\rho=1$, this means that
$\lambda_{\ast0}=0.033\cs k_1$; see \Eq{lamast0}.
This value will also be used to characterize the rotation rate
of the DNS, which are well characterized by the parameter
$\beta_\ast=0.33$; see \cite{BKKR12}.
Thus, $2\Omega/\lambda_{\ast0}$ is about 100 times larger than $\Co$.
Note also that in the MFS, $\Co=6\Omega\etat/\urms^2$
\citep{Jabbari2014}, which results in the same estimate.
The actual growth rates obtained from our DNS and MFS will be
normalized either also by $\lambda_{\ast0}$ or by $\tautd^{-1}$.

Both DNS and MFS simulations are done with the
\PCn\footnote{\tt{http://github.com/pencil-code}}.
It uses sixth order accurate finite differences and a third-order
timestepping scheme.
It comes with a special mean-field module that can be invoked for the
MFS.

\begin{table*}[t!]\caption{
Summary of run with different resolutions.
}\vspace{10pt}
\centerline{
\begin{tabular}{lccccccccccccc}
Run &  Resolution  & $B_0/\Beqz$
& $\Bfm/\Beqz$
&$\tautd^{\meanB_z=\max}$
&$(E_{\rm M}^{z,\max}/\Beqz^2)^{1/2}$
&$\tautd^{E_{\rm M}^z=\max}$
& $\Rey_{grid}$ &$\lambda\tautd$ &BR\\
\hline
\hline
A1 & $192^2\times384$  & 0.027 & 0.37 &1.2&0.06 (0.08) & 0.61 (1.6) & 0.018 &$0.74 \pm 0.08$ & yes \\%
A2 & $384^2\times768$  & 0.026 & 0.35 &0.9&0.08 (0.11) & 0.92 (0.9) & 0.018 &$1.28 \pm 0.13$ & yes \\%
A3 & $576^2\times1152$ & 0.025 & 0.42 &1.2&0.07 (0.10) & 0.61 (1.2) & 0.074 &$1.27 \pm 0.11$ & yes \\%
A4 & $768^2\times1536$ & 0.025 & 0.54 &1.5&0.13 (0.17) & 1.41 (1.4) & 0.001 &$1.27 \pm 0.14$ & yes \\%
A5 & $1152^2\times2304$& 0.025 & 0.40 &1.0&0.09 (0.12) & 0.92 (0.9) & 0.004 &$1.26 \pm 0.10$ & yes \\%
\hline
\hline
\label{runs1}
\end{tabular}}
\tablefoot{$\Bfm$ is the maximum of the Fourier-filtered vertical
  magnetic field strength at the surface ($z=0$),
  $\tautd^{\meanB_z=\max}$ is the time when $\Bfm/\Beqz$ reaches a
  maximum while $\tautd^{E_{\rm M}^z=\max}$ is the time when
$E_{\rm M}^z(k_\ast)$ with $k_\ast/k_1=2$ reaches a maximum.
The numbers in parentheses refer to spectral values averaged between
$k/k_1=1$ and $4$.
$\Rey_{\rm grid}= \urms / \nu k_{\rm Ny}$ is the mesh Reynolds number,
  $\Peffmin$ is the minimum value of the effective magnetic pressure,
  defined in \Sec{EffectiveMagneticPressure}.
The column BR indicates the visual appearance of BRs at the surface.
}
\end{table*}

\section{DNS results}

\subsection{Numerical resolution}
\label{Resol}

\begin{figure}[t!]\begin{center}
\includegraphics[width=\columnwidth]{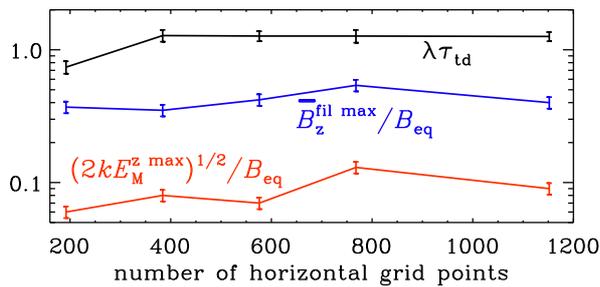}
\end{center}\caption[]{
Growth rate $\lambda$ (black line), the peak spectral
magnetic field $(2k_\ast E_{\rm M}^{z,\max})^{1/2}$ (red), and the maximum of
the Fourier-filtered vertical magnetic field $\Bfm$ (blue) for Runs~A1-A5 at
different resolutions.
}\label{growth_vs_res}\end{figure}

Since NEMPI is a mean-field instability, which relies on small-scale
turbulence for developing large-scale structures, we begin by
demonstrating the effects of changing the resolution on the formation
of magnetic field concentrations.
The growth rates are shown in \Fig{growth_vs_res} as a function
of resolution.
The corresponding simulations are listed in \Tab{runs1} where we have
always used the same domain size of $(2\pi)^2\times3\pi$, but with
different numbers of meshpoints.
We also quote the mesh Reynolds number,
$\Rey_{\rm grid}= \urms / \nu k_{\rm Ny}$, where
$k_{\rm Ny}=\pi/\delta x$ is the Nyquist wavenumber
and $\delta x$ is the mesh spacing.
In all cases, this number is well below unity.
Sometimes the quantity $\urms \delta x/ \nu$ is used in the
literature; it is simply $2\pi$ times larger than our $\Rey_{\rm grid}$.

\begin{figure}[t!]\begin{center}
\includegraphics[width=0.9\columnwidth]{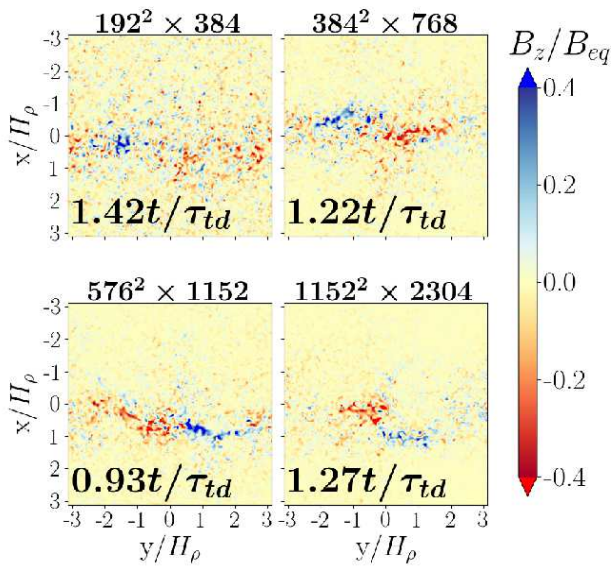}
\end{center}\caption[]{
Vertical magnetic field $B_z$ at the surface for runs with
different resolution (Run~A1 to Run~A3 and A5)
at the time when the structures are strongest.
}\label{fig:xy_resolution}\end{figure}

\begin{figure}[t!]\begin{center}
\includegraphics[width=\columnwidth]{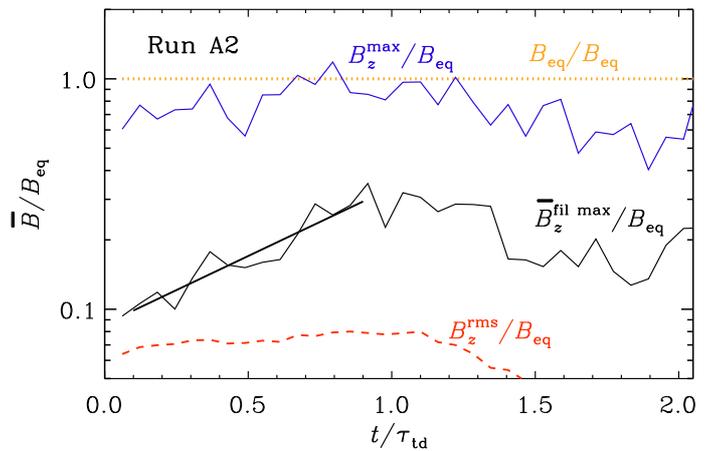}
\end{center}\caption[]{
Evolution of $\meanB_z^{\max}/\Beq$ vs.\ $t/\tautd$ for Run~A2
showing exponential growth with growth rate $\lambda=1.3\tautd^{-1}$
(black solid line), compared with that of $B_z^{\rm rms}/\Beq$
(red dashed line) and $B_z^{\max}/\Beq$ (blue solid line) at
$z=0$.
We also show the unity line corresponding to $\Beq$ (orange dotted line).
}\label{meanb_vs_t_growth}\end{figure}

\begin{figure}[t!]\begin{center}
\includegraphics[width=\columnwidth]{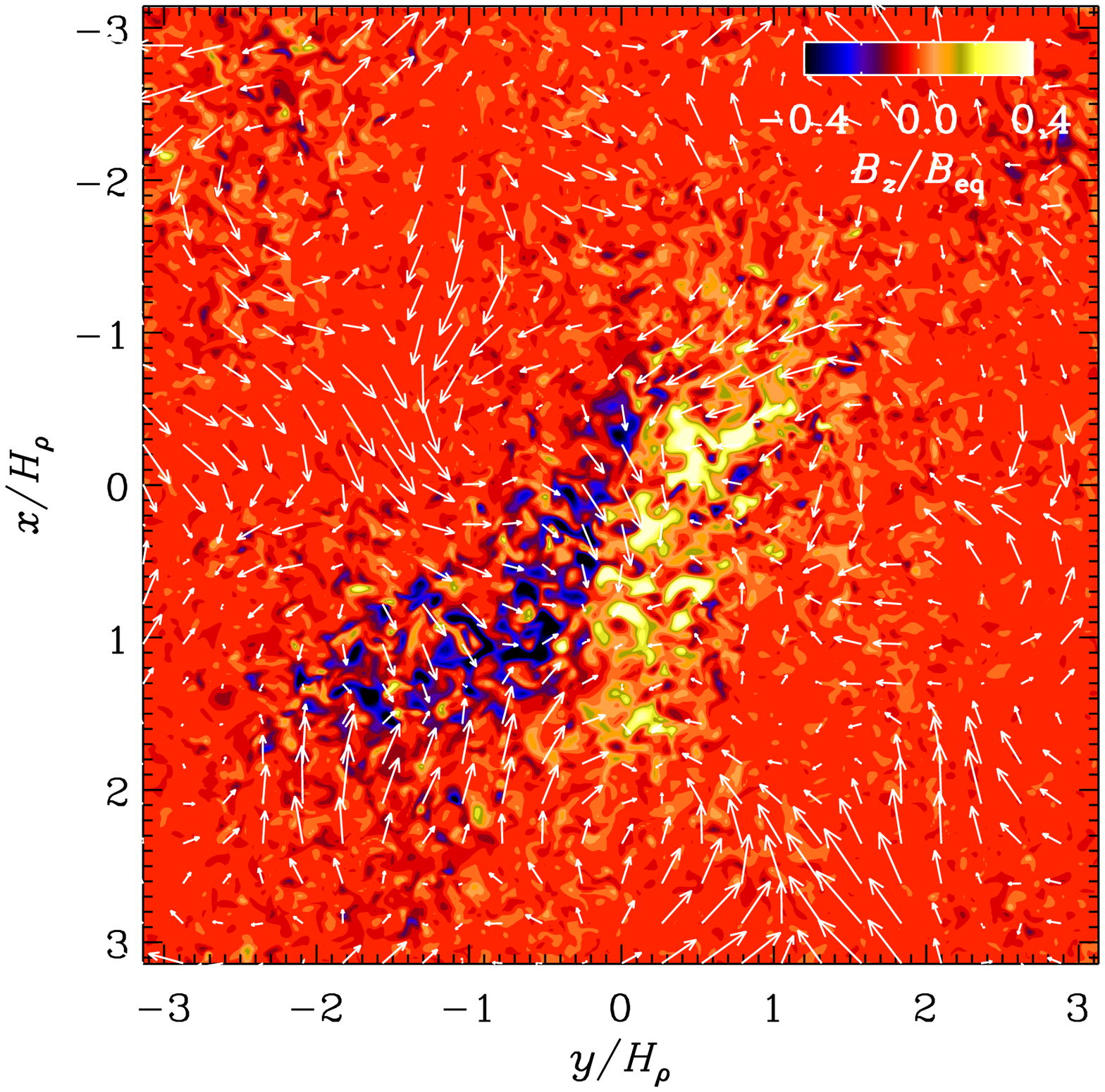}
\includegraphics[width=\columnwidth]{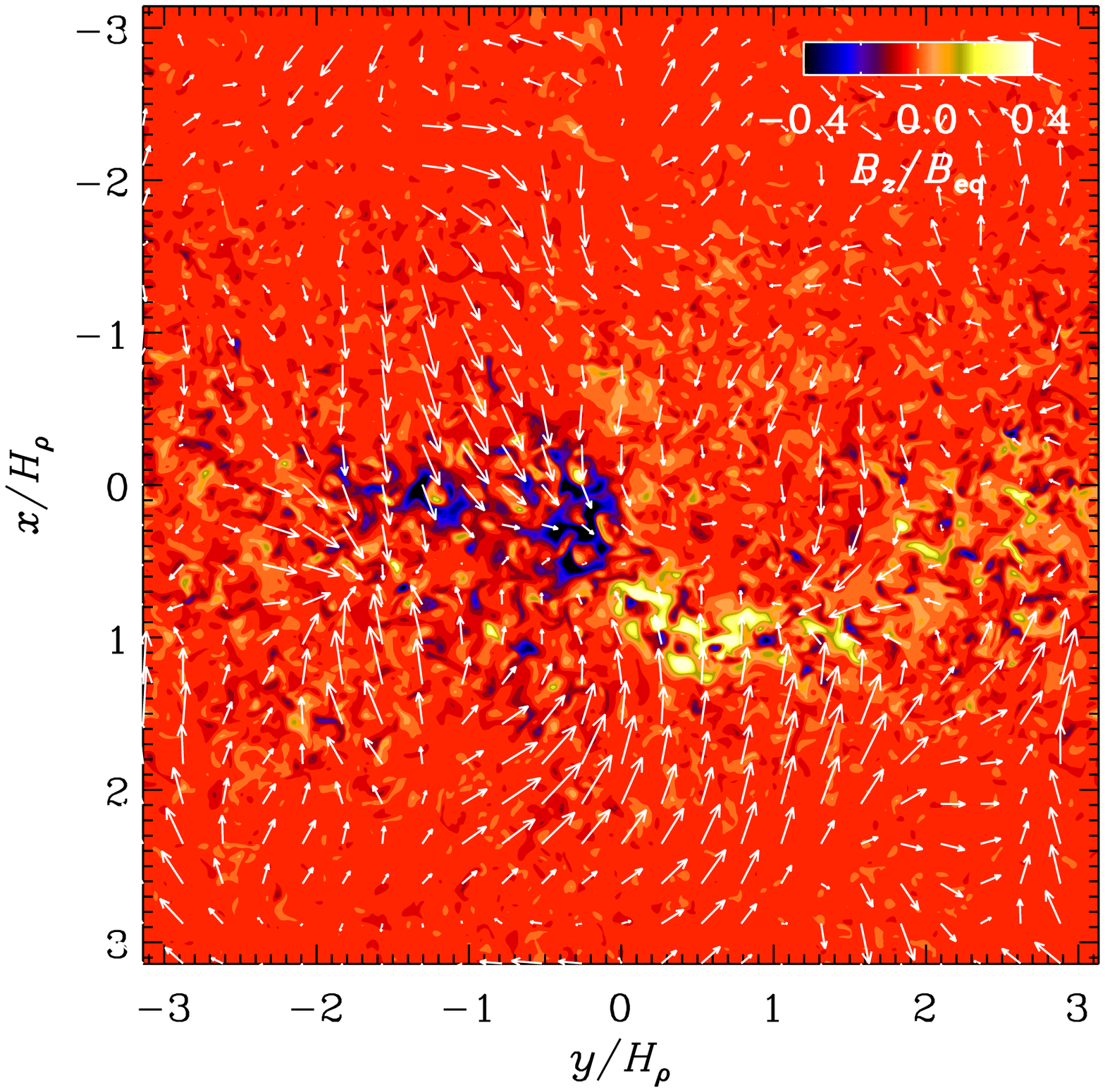}
\end{center}\caption[]{
Vectors of the mean (Fourier-filtered) flow superimposed
on a color scale representation of $B_z$ at the surface at
$t/\tautd=0.91$  for Run~A4 (upper panel) and
$t/\tautd=1.12$ for Run~A5 (lower panel).
Note that the coordinate system has been rotated by $90\degr$,
so $y$ points to the right and $X$ points downward.
}\label{prslice_single}\end{figure}

We see formation of bipolar regions (denoted by BR in the table) in all
the cases, but at the lowest resolution,
the growth rate of the magnetic field is significantly
smaller than at all higher resolutions.
Based on this, we conclude that a resolution of $384^2\times768$ meshpoints
results in a good compromise between accuracy and computational cost.
Therefore, we use simulations with this resolution for the following parameter study.
Because of this reason, \cite{WLBKR2015} doubled their resolution in
their followup work in \cite{WLBKR2013}.
Qualitatively, the coherence increases with increasing
resolution; see \Fig{fig:xy_resolution}.
We see from \Tab{runs1} the structures become strongest at a resolution
of $768^2\times1536$ meshpoints, where the normalized Fourier-filtered vertical
magnetic field at the surface, $\Bfm/\Beqz$, reaches
a value of 0.54 corresponding to only 0.4 at both lower and higher
resolutions.

\subsection{Growth of BRs}

We now discuss the main properties of BRs.
Large-scale BRs form during the first one or two
turbulent-diffusive times.
They are referred to as large-scale structures, because their size
is that of many turbulent eddies (about $2\pi/\kf$ in the horizontal
plane); see \Fig{prslice_single}.
To average over these turbulent eddies and still resolve the
large-scale structure of the BRs, we Fourier-filter the magnetic field
at the surface. To investigate the growth of these structures we then
plot the maximum of the vertical Fourier-filtered magnetic field
$\meanf{B}_z$ over time.
In \Fig{meanb_vs_t_growth}, which corresponds to
Run~A2, we clearly see an exponential growth with growth rate
$\lambda\approx1.3\,\tautd^{-1}$.
A similar growth has been seen before in numerical experiments both
with a horizontally imposed field \citep{BKKMR11, Kemel12b} and a vertical one
\citep{BKR2013}.
A similar value has also been determined with the same setup, where
BRs form \citep{WLBKR2015}.
Such exponential growth is suggestive of NEMPI.

In \Fig{meanb_vs_t_growth}, we also show for comparison the evolution
of the maximum of the surface vertical magnetic field $B^{\rm max}_z$,
i.e., not the filtered value.
Its value is always close to the local equipartition field strength.
No exponential growth phase can be seen in the temporal evolution
of $B^{\rm max}_z$.
Likewise, the rms value of surface vertical magnetic field
$B_z^{\rm rms} =\bra{B^2_z(z=0)}_{xy}^{1/2}$ is about 20 times smaller
and shows no exponential growth.

The formation of BRs in our simulations are associated with large-scale
flows.
As for the magnetic field, we perform Fourier filtering to
averaged over the turbulent scales.
In \Fig{prslice_single} we show these Fourier-filtered large-scale
flows that includes only wavenumbers below $\kf/6$.
The two panels are for Runs A4 and A5
with two different resolutions, but otherwise the same as
Run~A2, and similar times.
In both cases there are BRs, but in one case the two spots
are more separated from each other.
Nevertheless, in all cases the BRs are surrounded by a
large-scale inflow with relative rms value
$\meanU_{\rm hor}^{\rm rms}/\urms=0.16$--$0.18$,
where $\meanU_{\rm hor}^{\rm rms}=\bra{U_x^2+U_y^2}_{xy}^{1/2}$.
Its maximum value is around $0.4$.
Given that the scale of the turbulent motions
is much smaller than the scale of the spots, and that there is otherwise no
mechanism producing large-scale flow perturbations,
the inflow can only
be a consequence of the magnetic field itself, and not the other way around.
This is indeed also what one expects for NEMPI.

\begin{figure}[t!]
\centering
\includegraphics[width=\columnwidth]{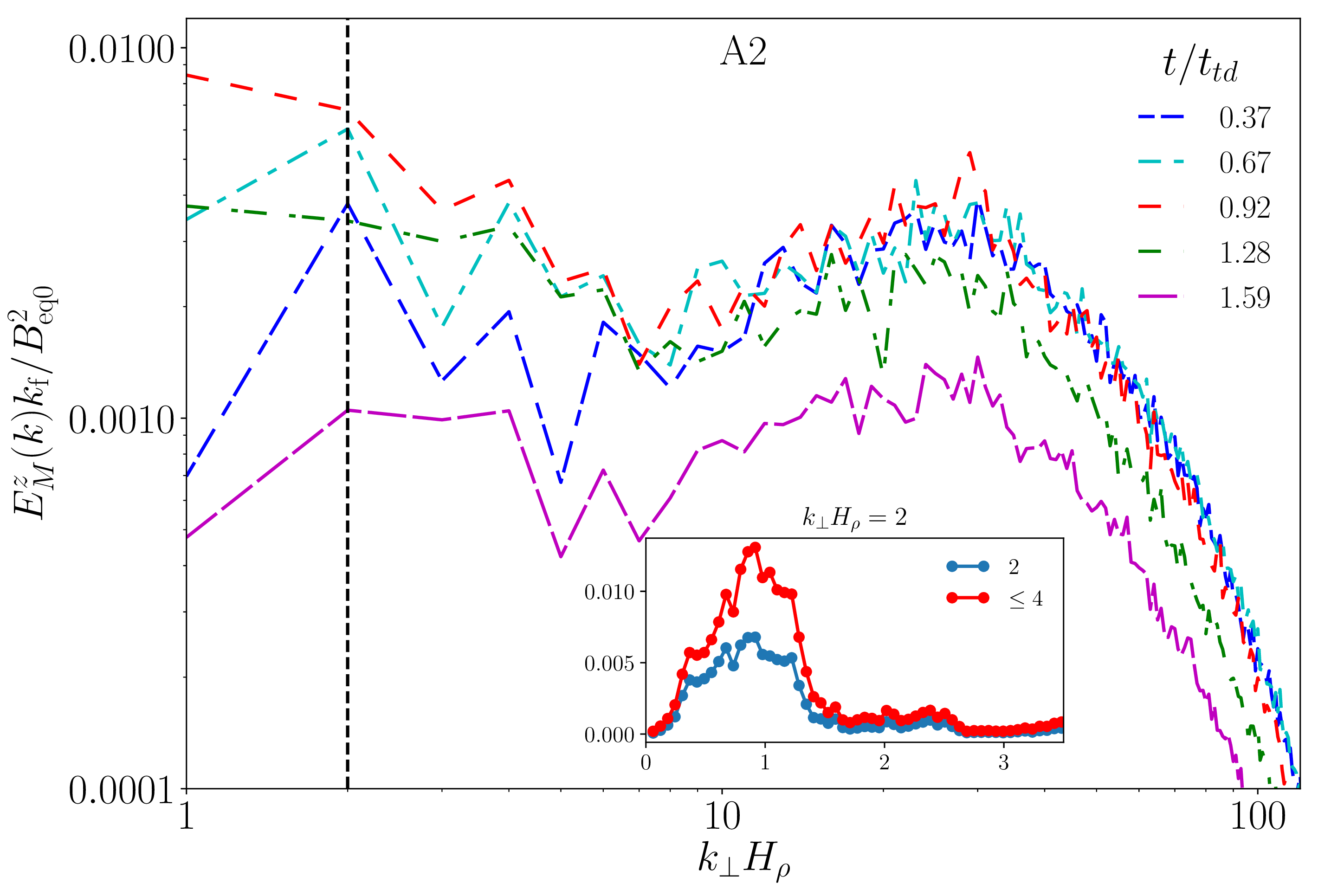}
\label{fig:pspecb}
\caption{Power spectra of $B_z$ at the surface ($z=0$) for Run~A2 at
several
times around the maximum growth for zero rotation.
The inset shows the time dependence (time in units of $\tautd$)
for $k=k_\ast=2\,k_1$ and for the values averaged between $k/k_1=1$ and $4$.
}\label{pspec}
\end{figure}

Another diagnostics for the formation of magnetic structures in
our simulations are magnetic power spectra taken at the surface.
In \Fig{pspec} we can see the time evolution of such power spectra.
At early times when there are no structures yet, the spectrum peaks at
the energy injection wavenumber,
$\kf=30\,k_1$.
As time evolves and structures start forming near the surface,
magnetic energy is transported toward smaller wavenumbers.
When we see the BRs forming at the surface, the magnetic
power spectrum peaks at $k=k_\ast\approx2k_1$, and
the amplitude at this wavenumber decreases until the structures disappear.
The strength of the magnetic surface structures at this wavenumber is characterized
by the value of $(2kE_{\rm M}^{z,\max})^{1/2}$ at $k=k_\ast=2\,k_1$; see
\Sec{Sec:DNS}.
The resulting values are listed in \Tabss{runs1}{runs3}.
In addition, we also give the spectral values averaged over the first
four wavenumbers from $k/k_1=1$ to $4$.
The averaging helps reducing the sensitivity to discretization noise
that arises from looking at just one wavenumber $k_\ast$.
In the tables, we also judge BR formation as yes, no, or weak.
These attributes are based on the qualitative assessment of images
of $B_z$.

\begin{table*}[t!]\caption{
Summary of runs with different Coriolis numbers and colatitudes.
}\vspace{10pt}
\centerline{
\begin{tabular}{lccccccccc}
Run &   $\Co$ & $\theta$
   & $\Bfm/\Beqz$
   &$\tautd^{\meanB_z=\max}$
   &$\sqrt{2k_\ast E_{\rm M}^{z,\max}}/\Beqz$
   &$\tautd^{E_{\rm M}^z=\max}$
   &$\lambda\tautd$ &$\Peffmin$&BR\\
\hline
\hline
A2 & 0 & 0 &  0.35 &0.9&0.08 (0.11) & 0.92 (0.9) & $1.28 \pm 0.13$ &$-0.029$& yes \\%
\hline
B1 & 0.0012 & 0  & 0.31 & 0.8 &0.17 (0.22) &0.61 (0.61)&$1.32\pm 0.20$ &$-0.029$& yes \\
B2 & 0.0012 & 30 & 0.35 & 0.9 &0.13 (0.18) &0.61 (0.73)&$1.37\pm 0.11$ & & yes \\
B3 & 0.0012 & 60 & 0.32 & 1.7 &0.16 (0.20) &0.67 (0.67)&$1.46\pm 0.16$ & & yes \\
B4 & 0.0012 & 90 & 0.42 & 1.8 &0.13 (0.16) &0.73 (0.67)&$0.58\pm 0.06$ & & yes
\\
\hline
C1 &  0.0023 & 0 & 0.672 & 1.7 &0.30 (0.42) &1.59 (1.59)& $1.03\pm 0.07$ &$-0.025$& yes\\
C2 &  0.0023 & 30& 0.303 & 0.8 &0.12 (0.16) &0.61 (0.61)& $1.46\pm 0.19$ &$-0.021$& yes\\
C3 &  0.0023 & 60& 0.428 & 1.7 &0.18 (0.26) &1.65 (1.65)& $0.66\pm 0.06$ &$-0.019$& yes\\
C4 &  0.0023 & 70& 0.341 & 2.4 &0.12 (0.16) &3.18 (1.96)& $0.24\pm 0.05$ & & yes\\
C5 &  0.0023 & 90& 0.467 & 2.0 &0.14 (0.20) &1.16 (1.96)& $0.64\pm 0.06$ &$-0.020$& yes\\
\hline
D1 &  0.0029 & 0 & 0.532 & 1.41 &0.26 (0.38) & 1.16 (1.28)& $1.40\pm 0.17$ &$-0.028$& yes\\
D2 &  0.0029 & 60& 0.406 & 2.63 &0.16 (0.22) & 1.16 (1.96)& $0.37\pm 0.05$ & & yes\\
\hline
E1 & 0.0035 & 0 & 0.228 & 0.6 &0.11 (0.15) & 0.9 (0.9) &$0.92\pm 0.3$ &$-0.023$& yes\\
E2 & 0.0035 & 60& 0.241 & 1.1 &0.13 (0.16) & 0.5 (0.5) &$0.56\pm 0.1$ & & weak \\
\hline
F1 &  0.0076 & 0 & 0.296 & 1.4&0.11 (0.14) & 1.5 (1.5)& $0.38\pm 0.09$ &$-0.029$& weak\\
F2 &  0.0076 & 30& 0.269 & 3.0&0.12 (0.16) & 1.8 (1.8)& $0.17\pm 0.05$ & & yes\\
F3 &  0.0076 & 60& 0.226 & 0.5&0.11 (0.14) & 0.3 (0.3)& $0.16\pm 0.13$ & & no\\
F4 &  0.0076 & 90& 0.209 & 0.7&0.11 (0.15) & 0.8 (0.8)& $0.60\pm 0.15$ & & weak
\\
\hline
\hline
\label{runs2}
\end{tabular}}
\tablefoot{
All runs have a resolution of $384^2\times768$ meshpoints (as in Run~A2),
an imposed field of $B_0/\Beqz \approx 0.026$,
and a size of $(2\pi)^2\times3\pi$.
The column BR indicates the visual appearance of BRs at the surface.
All other quantities are defined in \Tab{runs1}.
}\end{table*}

\begin{table*}[t!]\caption{
Summary of runs with different values of imposed field
and Coriolis number.
}\vspace{10pt}
\centerline{
\begin{tabular}{lccccccccc}
Run &   $B_0/\Beqz$& $\Co$
   & $\Bfm/\Beqz$
   &$\tautd^{\meanB_z=\max}$
   &$\sqrt{2k_\ast E_{\rm M}^{z,\max}}/\Beqz$
   &$\tautd^{E_{\rm M}^z=\max}$
   &$\lambda\tautd$ & $\Peffmin$ &BR\\
\hline
\hline
A2 & 0.026 & 0 &  0.35 &0.9&0.08 (0.11) & 0.92 (0.9)      &$1.28\pm 0.13$ &$-0.029$& yes \\%
\hline
G1 & 0.065 & 0.002 & 0.38 & 0.4& 0.32 (0.40)& 0.49 (0.55) &$1.21\pm 0.20$ &$-0.046$& yes \\
G2 & 0.066 & 0.004 & 0.54 & 0.9& 0.32 (0.46)& 0.73 (0.73) &$1.27\pm 0.12$ &$-0.041$& yes \\
G3 & 0.066 & 0.006 & 0.52 & 1.6& 0.38 (0.52)& 1.59 (1.59) &$0.67\pm 0.05$ &$-0.039$& yes \\
G4 & 0.067 & 0.012 & 0.50 & 2.5& 0.38 (0.50)& 1.34 (1.34) &$0.50\pm 0.06$ &$-0.048$& yes \\
G5 & 0.066 & 0.015 & 0.46 & 1.3& 0.32 (0.42)& 0.98 (1.89) &$0.34\pm 0.07$ & & weak \\
G6 & 0.068 & 0.018 & 0.43 & 1.4& 0.28 (0.34)& 0.98 (0.98) &$0.33\pm 0.06$ &$-0.036$& no \\
\hline
H1 & 0.13  & 0.002 & 0.50& 0.8&0.28 (0.38) &0.5 (0.3)& $0.55\pm 0.09$ &$-0.069$& yes\\ 
H2 & 0.14  & 0.006 & 0.49& 0.5&0.32 (0.42) &0.5 (0.5)& $0.22\pm 0.13$ &$-0.070$& yes\\ 
H3 & 0.14  & 0.012 & 0.63& 2.3&0.44 (0.64) &0.7 (2.1)& $0.20\pm 0.03$ &$-0.067$& yes\\ 
H4 & 0.14  & 0.018 & 0.54& 3.0&0.32 (0.54) &3.5 (0.9)& $0.09\pm 0.04$ &$-0.068$& weak\\
\hline
I1 & 0.27 & 0.002 & 0.59 & 0.4 &0.24 (0.34)&0.4 (0.4)& $0.50\pm 0.52$ &$-0.12$& yes\\ 
I2 & 0.28 & 0.006 & 0.61 & 0.4 &0.44 (0.58)&0.4 (0.4)& $0.56\pm 0.29$ &$-0.12$& yes\\
I3 & 0.29 & 0.013 & 0.56 & 1.0 &0.42 (0.64)&0.6 (0.5)& $0.21\pm 0.08$ &$-0.12$& weak\\ 
I4 & 0.30 & 0.019 & 0.59 & 1.0 &0.42 (0.74)&0.7 (1.0)& $0.15\pm 0.07$ &$-0.12$& very weak\\ 
\hline
J1 & 0.77 & 0.003 & 0.69 & 0.05&0.20 (0.28) &0.22 (0.22)& $0.10\pm 0.29$ &$-0.15$& very weak\\ 
J2 & 0.82 & 0.007 & 0.64 & 0.21&0.21 (0.26) &0.32 (0.32)& $0.05\pm 0.10$ &$-0.16$& very weak \\
J3 & 0.91 & 0.014 & 0.69 & 0.37&0.23 (0.30) &1.31 (1.31)& $0.02\pm 0.12$ &$-0.15$& no\\
J4 & 0.91 & 0.021 & 0.73 & 0.73&0.31 (0.38) &1.36 (1.36)& $0.03\pm 0.10$ &$-0.16$& no\\
\hline
\hline
\label{runs3}
\end{tabular}}
\tablefoot{
The colatitude is set to $\theta=0\degr$, corresponding to the pole.
All other quantities are defined in \Tabs{runs1}{runs2}.
}\end{table*}

\begin{figure*}[t!]\begin{center}
\includegraphics[width=0.9\textwidth]{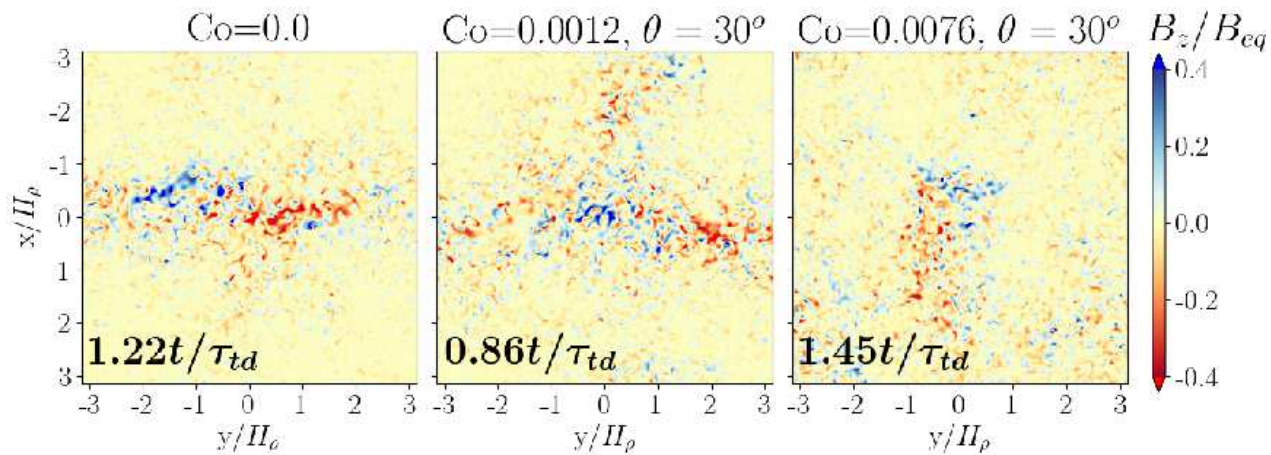}
\end{center}\caption[]{
Rotational dependency on the active region formation.
We show the vertical magnetic field $B_z/\Beq$ for Runs~A2, B2 and F2
corresponding to $\Co = 0$, $0.0012$, $0.0076$ at the time of the strongest BR.
In the first panel for $\Co = 0$, the value of $\theta$ is insignificant.
}\label{bb3_xy}
\end{figure*}

The energy transfer to larger scales
is reminiscent of an inverse cascade.
\cite{BGJKR14} have speculated that such a cascade might be a
consequence of the conservation of cross helicity, $\overline{\uu\cdot\bb}$,
where the overbar denotes horizontal averaging,
$\bb=\BB-\meanBB$ are the magnetic fluctuations and
$\uu=\UU-\meanUU$ are the velocity fluctuations;
see \cite{RKB11}.
(For horizontal averages, we usually have $\meanUU=0$.)
They studied the production of $\overline{\uu\cdot\bb}$ as a result of a mean
magnetic field along the direction of gravity, so there exists a
pseudoscalar $\grav\cdot\BB_0$ that has the same symmetry properties as
$\overline{\uu\cdot\bb}$ and is also odd in the magnetic field.
A similar result was also obtained by \cite{KKMRSZ03}; see their
Eq.~(11) where they considered inhomogeneous turbulence.
Furthermore, recent work of \cite{ZB18} has shown that the cross
helicity spectrum shows a steep slope at large scales.
This was interpreted as a potential signature of NEMPI-like effects.

\subsection{Influence of rotation}
\label{InfluenceRotation}

The structure of the bipolar regions is strongly influenced by
rotation; see \Fig{bb3_xy}.
Faster rotation causes the magnetic flux concentrations to be
weaker. In some cases, for example in Run~B3
(middle panel of \Fig{bb3_xy}),
the structure of BRs splits into three parts with one
negative and two positive polarities.
In our most rapidly rotating case (right-most panel of \Fig{bb3_xy}), the
structure appears rotated by $90\degr$ with respect to slower rotations.

\Tab{runs2} shows that, as the Coriolis number is increased from 0.0012 to
0.0076, BR formation gets almost entirely suppressed (see Runs~B1 to F4).
At $\theta=0\degr$ (corresponding to the pole), BR formation tends
to be slightly easier; but as the Coriolis number is increased,
the BRs become clearer at intermediate latitudes.

\begin{figure}[t!]\begin{center}
\includegraphics[width=\columnwidth]{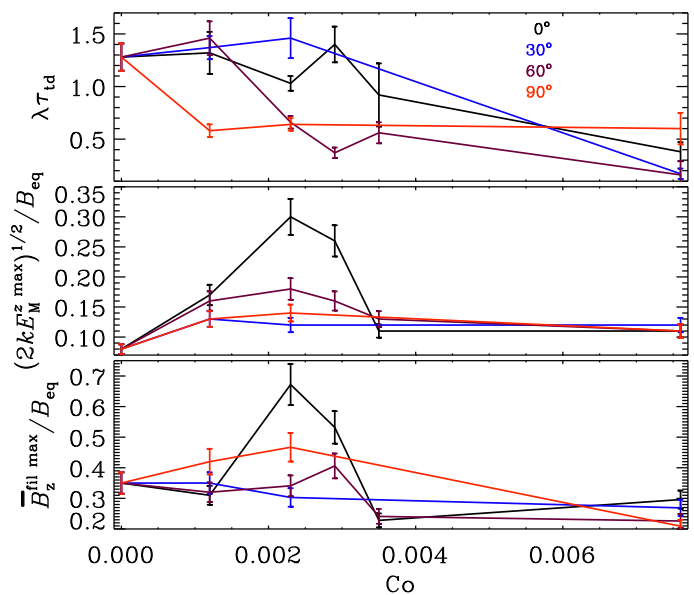}
\end{center}\caption[]{
Rotational dependency of the growth rate $\lambda$, the peak spectral
magnetic field $(2k_\ast E_{\rm M}^{z,\max})^{1/2}$, and the maximum of
the Fourier-filtered vertical magnetic field $\Bfm$ for colatitudes
$0^\circ$ (black line), $30^\circ$ (blue line), $60^\circ$ (purple
line), and $90^\circ$ (red line).
The error bars are either the errors of the exponential fit
($\lambda$; see \Tab{runs2}), or estimated as $10\%$ of the actual
value.
}\label{pLam_co}\end{figure}

The rotational dependency of the growth rate $\lambda$
for different colatitudes $\theta$ is shown in \Fig{pLam_co}.
For most of the colatitudes, the growth rate first increases for weak
rotation and then gets reduced to around a third of the values for more
rapid rotation.
Even though we cannot visually detect any clear indication of BRs in
the rapidly rotating runs,
the growth rate of magnetic field is still positive.
The magnetic field strength determined from the spectral energy (see
middle panel of \Fig{pLam_co}) shows
actually an enhancement with
rotation. Even for the rapidly rotating cases, the magnetic field
strength in the flux concentrations
is stronger than without any rotation. The strongest values
are achieved with $\Co=0.002$.
For most of the colatitudes, the maxima of the large-scale magnetic
field, as shown in the bottom row of \Fig{pLam_co}, decrease for increasing
rotation, similarly as the $\lambda$.
However, we see the maximum value for $\Co=0.002$ to $0.003$.

The time when the Fourier-filtered magnetic field
and the field of spectral
energy become maximal depends on rotation, but there is no clear
trend visible; see \Tabs{runs2}{runs3}. However there seems to be an indication
that the time becomes longer, as expected for a smaller growth rate.
As the Coriolis number increases, the structures become weaker and it gets more
difficult to discern a clear rotation pattern.

We note here that, unlike in convection, the energy-carrying length scale is
not influenced by rotation, because the driving scale is prescribed through
the forcing function. Similarly, the kinetic energy is also only weakly
influenced by rotation, so $\urms$ decreases only weakly for more rapid
rotation.

\subsection{Dependence on latitude}

\begin{figure}[t!]\begin{center}
\includegraphics[width=\columnwidth]{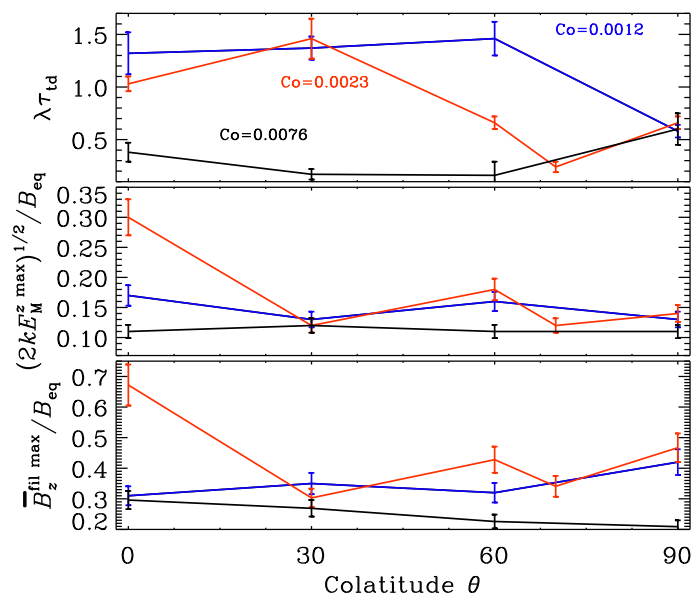}
\end{center}\caption[]{
Summary of the results for the different colatitudes showing the normalized
growth rate of the Fourier-filtered $z$-component of the magnetic field
(top panel) for $B_0/\Beqz = 0.02$,
the averaged magnetic spectrum over $k\leq 4$ (middle panel) and the
maximum of the Fourier-filtered vertical magnetic field $\Bfm$ (bottom panel)
for different values of the Coriolis number and colatitude.
As elsewhere, error bars are either the errors of the exponential fit
($\lambda$; see \Tab{runs3}), or estimated as $10\%$ of the actual
value.
}\label{bzVStheta}\end{figure}

As we change the colatitude $\theta$, the growth rates and the
strengths of BRs change.
This is demonstrated in \Fig{bzVStheta}, where
we show $\lambda\tautd$, $(2k_\ast E_{\rm M}^{z,\max})^{1/2}/\Beqz$ and
$\Bfm/\Beqz$
for $\Co=0.0012$ (corresponding to
$2\Omega/\lambda_{\ast0}\approx0.15$)
and different values of $\theta$.
For $\theta=0$, which corresponds to the pole,
the saturation magnetic field strength shows a maximum
at $t/\tautd\approx1$.
For larger values of $\theta$, i.e., closer to the equator,
the maximum is slightly higher (about 0.8) and more long-lived,
e.g., for $0.5\leq t/\tautd\leq2.5$ at $\theta=60\degr$, when
$\meanB_z^{\rm rms}/\Beq$ is above 0.6 and sometimes even 0.8.

\begin{figure}[t!]\begin{center}
\includegraphics[width=0.9\columnwidth]{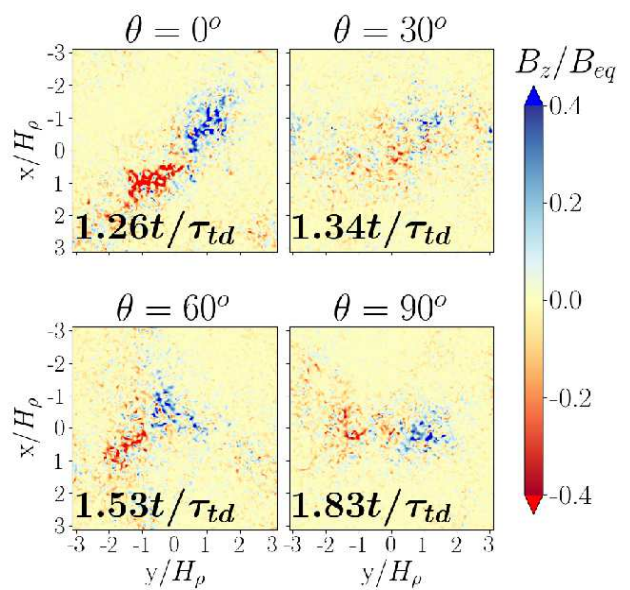}
\end{center}\caption[]{
Vertical surface magnetic field $B_z/\Beq$ for Runs~C1--C3 and C5 with $\Co = 0.0012$.
}\label{bb3_xy01}
\end{figure}

Bipolar structures are still fairly pronounced at $\theta=60\degr$,
i.e., at $30\degr$ latitude; see \Fig{bb3_xy01}.
It is remarkable that for all values of $\theta$,
the inclination of BRs is approximately the same.
The same feature is also seen in the MFS, except that for
$\theta=90\degr$ the structures are always aligned in the $y$ direction.
This may here be the case, too, but the structures are so weak
that this is hard to see.
For somewhat faster rotation, when $\Co>0.01$, corresponding to
$2\Omega/\lambda_{\ast0}\approx1$, the structures disappear.

\subsection{Inclination of BRs}
\label{Inclination}

A systematic orientation of BRs was already seen in the original
papers of \cite{LBKMR2012,LBKR2013a} both for MFS and DNS.
For most of the runs, the BRs are either aligned
with the imposed magnetic field or they are inclined by $45^\circ$.
To compare with the Sun, we map our Cartesian coordinate system
to spherical ones via $(x,y,z)\to(\theta,\phi,r)$, so the
$y$ coordinate points in the toroidal (eastward) direction and
$x$ corresponds to colatitude, which points southward.
This explains the orientations of our surface visualizations
where we plot $-x$ versus $y$; see \Fig{bb3_xy01}.

We usually find that rotation leads to a poleward tilt of the BR.
We return to the question of the inclination angle in
\Sec{InclinedSurfaceStructures} where we study a similar
phenomenon in MFS and show that the inclination angle is then not
an artifact of the domain size.
Note also that the orientation is the same in DNS and MFS.
In fact, the orientation agrees with that found by
\cite{LBKMR2012,LBKR2013a}.
It is interesting to note that the sense of inclination
is the other way around than what is expected based on the buoyant
rise of magnetic flux tubes, which gives rise to Joy's law \citep{CG87}.

The ``anti-Joy's'' law orientation of NEMPI structures is likely a
consequence of the interaction of rotation with the concentration of
flux as opposed to the expansion of flux, which is usually expected as
a flux tube rises through a stratified layer.
A similar phenomenon of a concentration of flux in stratified turbulence
(as opposed to an expansion) has been found in turbulence driven by the
magneto-rotational instability, where this was argued to be the reason
for an unconventional sign of the $\alpha$ effect \citep{BNST95}; see
the more detailed discussion of \cite{BC97}.

\subsection{Dependence on the imposed magnetic field strength}
\label{DependenceFieldStrength}

Rotation weakens the formation of structures for even smaller Coriolis
numbers than those of previous studies \citep{LBKMR2012,LBKR2013a}.
We can therefore increase the efficiency of the formation mechanism
by increasing the imposed magnetic field,
as was suggested by \cite{WLBKR2015}.
The increase in the imposed field
allows the BRs to be formed for even
higher Coriolis numbers, up to the point where
the magnetic field becomes so strong that
the derivative $\dd\Peff/\dd \mean{B}{}^2$ becomes positive and NEMPI cannot be
excited in the domain;
for details, see \cite{BRK2016} and the appendix of \cite{Kemel13a}.

\Tab{runs3} shows simulations of domains located at the pole ($\theta=0\degr$)
and gives their dependence on the magnetic field strength and angular velocity.
As $B_0/\Beqz$ is doubled from 0.07 to 0.14, BR formation becomes slightly
easier: compare model G6 with model H4 (both are for $\Co=0.018$).
Weak BR formation is only possible in model H4.
For even stronger fields, however, this trend disappears.
In model I4 with $B_0/\Beqz=0.3$, BR formation is now very weak and
for $\Co=0.019$, and in model J3 with $B_0/\Beqz=0.9$,
BR formation is impossible---even for $\Co=0.014$; see
\Tab{TCoB0Beq0} for an overview.

\begin{table}[b!]\caption{
Comparison of BR formation as a function of rotation in terms of Co
and imposed magnetic field $B_0/\Beqz$.
}\vspace{12pt}\centerline{\begin{tabular}{l|cccc}
      &     \multicolumn{4}{c}{$B_0/\Beqz$}  \\
\hline
Co    &  0.07 & 0.14  &  0.3  &  0.8  \\
\hline
\hline
0.0   &  yes  & yes   &       & no\\
0.002 &  yes  & yes   &  yes  &very weak\\
0.006 &  yes  & yes   &  yes  &very weak\\
0.012 &  yes  & yes   &  weak &   no   \\
0.018 &   no  & weak &very weak&  no   \\
\hline
\label{TCoB0Beq0}\end{tabular}}
\tablefoot{The values for $\Co$ were taken from \cite{WLBKR2015}.}
\end{table}

\begin{figure}[t!]\begin{center}
\includegraphics[width=\columnwidth]{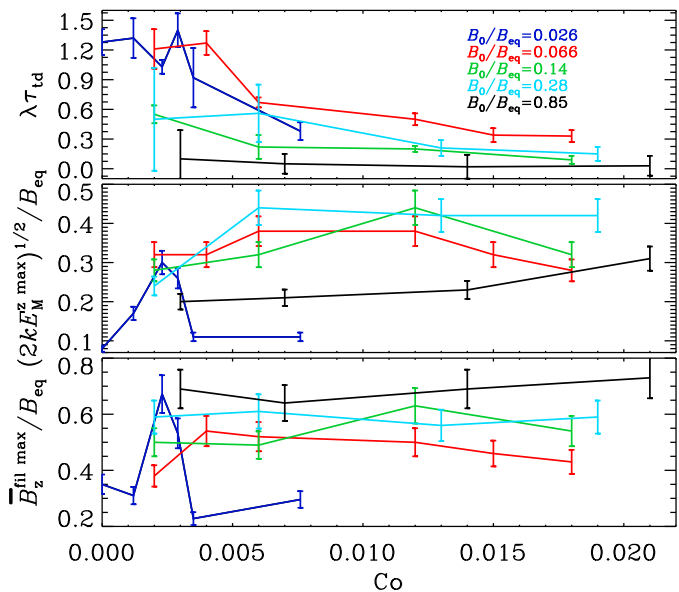}
\end{center}\caption[]{
Rotational dependency of the growth rate $\lambda$, the peak spectral
magnetic field $(2k_\ast E_{\rm M}^{z,\max})^{1/2}$, and the maximum of
the Fourier-filtered vertical magnetic field $\Bfm$ for various averaged
imposed magnetic field strengths $B_0/B_{\rm eq}=0.026$ (blue line),
$0.066$ (red), $0.14$ (green), $0.28$ (cyan), and $0.85$ (black).
Error bars denote either the errors of the exponential fit
($\lambda$; see \Tab{runs2}), or are estimated as $10\%$ of the actual
value.
}\label{pLam_B0}\end{figure}

The growth rate of the magnetic field
shows a strong decrease for higher rotation rates. This can
be compensated for to some extent by using a stronger imposed magnetic field; see the top
panel of \Fig{pLam_B0}.
For $\Co\approx0.006$, the growth rates with imposed magnetic fields
of $B_0/B_{\rm eq}=0.066$ and $B_0/B_{\rm eq}=0.28$ are indeed higher
than for $B_0/B_{\rm eq}=0.026$.
A similar behavior can be found by looking at the magnetic field
strength determined from the spectral energy (see middle panel of
\Fig{pLam_B0}). For imposed magnetic field strengths between $B_0/B_{\rm
eq}=0.066$ and $0.85$, the magnetic field is not
much influenced by rotation and stays roughly constant above
$\Co\approx0.006$ at a level that is even higher than for smaller
rotation.
The large-scale magnetic field $\Bfm$ for imposed magnetic fields
larger than $B_0/B_{\rm eq}=0.06$ does not show a strong rotational
influence. The large-scale magnetic field is higher for larger imposed
magnetic field and keeps this level for a large range of rotation rates.
This means that a higher imposed magnetic field can indeed prevent
NEMPI from being quenched for larger rotation rates. Even
with a large imposed magnetic field, rotational quenching takes place,
but at larger rotation rates.

\subsection{Effective magnetic pressure}
\label{EffectiveMagneticPressure}

\begin{figure}[t!]\begin{center}
\includegraphics[width=\columnwidth]{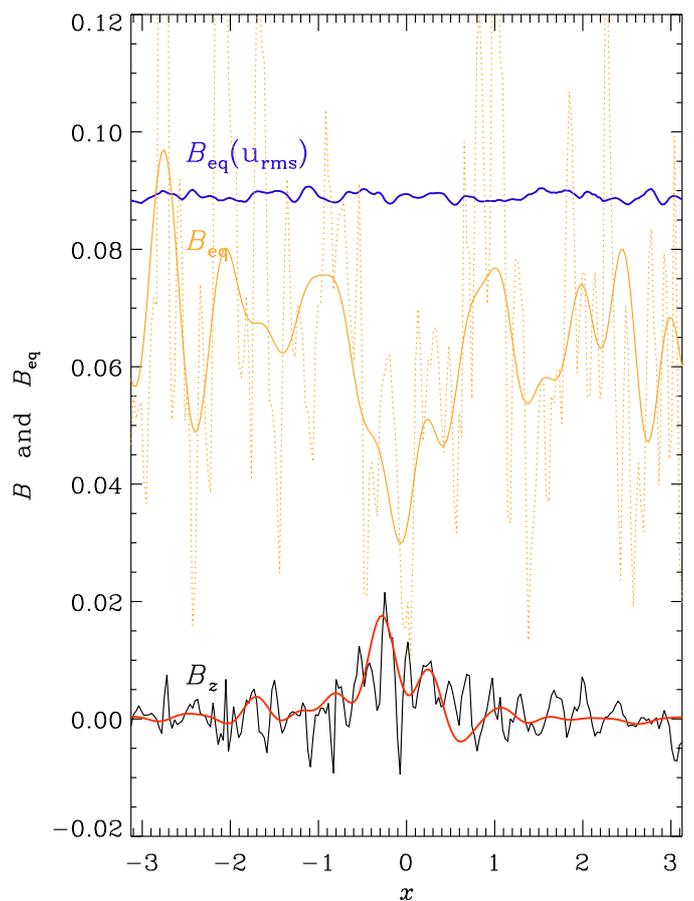}
\end{center}\caption[]{
Profiles of $\Beq$ and $B_z$ at $z=0$ (i.e., the surface of the turbulent
region), for Run~B1 with $\Co=0.0015$, $\theta=0$ at $t/\tautd=1$
and along $y=0$ (black line).
The red solid line gives the Fourier-filtered profiles of $B_z$.
The dotted orange line gives the values of $\Beq$ through the magnetic
structure and the solid orange lines denotes its Fourier-filtered value.
The blue line represents the value of $\Beq$ based on the volume-averaged
velocity, but the local density.
All values have been normalized by the volume-averaged value of $\Beq$.
}\label{b_beq22_Om002_theta0}\end{figure}

The concentration of magnetic field in a NEMPI scenario is possible
due to a turbulence effect on the effective magnetic
pressure, which is the sum of non-turbulent and turbulent contributions
to the large-scale magnetic pressure.
This effect results in a suppression
of the total (hydrodynamic plus magnetic)
turbulent pressure by the large-scale (mean) magnetic field.
This means that an increase of the mean magnetic field due to the instability
will be accompanied by a decrease of the turbulent pressure
and a reduction of the equipartition field strength, $\Beq$.
This is because the hydrodynamic part of the total turbulent pressure,
$p_{\rm hydro}=\meanrho\urms^2/3=\Beq^2/3$,
as well as the turbulent kinetic energy density,
decrease due to an increase of
the turbulent magnetic energy density as well as
the turbulent magnetic pressure
through tangling magnetic fluctuations \citep{KMR96,RK07}.

\FFig{b_beq22_Om002_theta0} shows the profiles of $\Beq$ and $B_z$
at the surface along $y=0$ at $t/\tautd=1$.
This figure clearly shows that at the location of the maximum of the
vertical magnetic field, the equipartition field $\Beq$ is decreased.
Most of this suppression comes from a local decrease in the turbulent
velocity, while the local density through the structure varies only
little; see the blue line of \Fig{b_beq22_Om002_theta0}.
We attribute this behavior to the operation of NEMPI in the
simulation.

\begin{figure}[t!]\begin{center}
\includegraphics[width=\columnwidth]{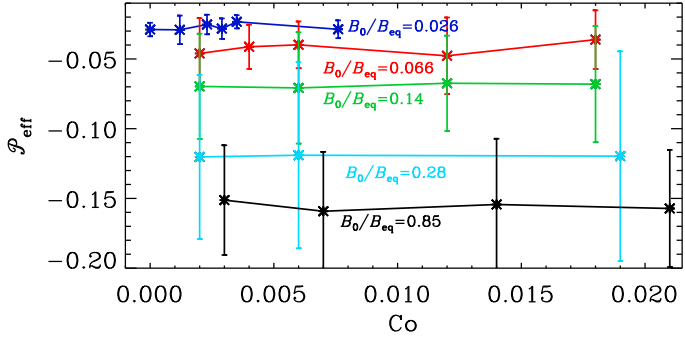}
\end{center}\caption[]{
Dependence of the minimum value of the normalized effective magnetic
pressure, $\Peff$, on the Coriolis number for different values of the
imposed field strength, $B_0/\Beqz$ (the curves have been labeled
by an average value), for the runs listed in \Tabs{runs2}{runs3}.
Again, error bars are estimated using the maximum difference of the total mean
with the means of each third of the time series.
}\label{peffmin}
\end{figure}

\begin{figure}[t!]\begin{center}
\includegraphics[width=\columnwidth]{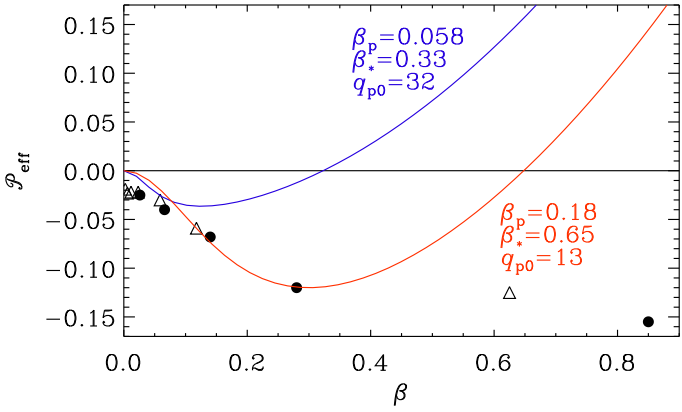}
\end{center}\caption[]{
Dependence of $\Peff(\beta)$ obtained from \Fig{peffmin} by taking an
average for different values of $\Co$ (filled circles).
The blue line is our standard representation for
$\qpz=32$ and $\betap=0.058$, which corresponds to $\beta_\ast=0.33$,
while the red line is a better fit to the present data giving
$\qpz=13$ and $\betap=0.18$, which corresponds to $\beta_\ast=0.65$.
Note that the data point at $\beta=0.85$ and $\Peff=-0.16$ is well
outside any $\Peff(\beta)$ curve that could fit all the data and
has therefore been discarded as an ``outlier''.
As a comparison, we plot the values of \cite{WLBKR2015} as triangles.
}\label{ppeff}
\end{figure}

Next, we determine the normalized effective magnetic pressure,
$\Peff$, as
\begin{equation}
\Peff=(1-\qp)\,\beta^2/2,
\label{peff}
\end{equation}
where $\qp$ is obtained from \citep{Kemel13a}
\begin{equation}
\qp=-{1\over\mean{\BB}^2} \,
\left(\Delta\overline\Pi_{xx}+\Delta\overline\Pi_{yy}
-\left(\Delta\overline\Pi_{xx}-\Delta\overline\Pi_{yy}\right)\,{\mean{B_x}^2
+\mean{B_y}^2\over\mean{B_x}^2 -\mean{B_y}^2} \right).
\label{q_p}
\end{equation}
Overbars denote here $xy$ averaging and the diagonal components of the
total (Reynolds and Maxwell) stress
tensor, $\Delta\overline\Pi_{ii}$, have been obtained from the DNS as
\EQ
\Delta\overline\Pi_{ii}
=\meanrho\,(\overline{u_i^2}-\overline{u_{0i}^2})
+\half(\overline{\bb^2}-\overline{\bb_0^2})
-(\overline{b_i^2}-\overline{b_{0i}^2}),
\EN
where the subscript $0$ refers to values for $B_0=0$
and lower case letters denote fluctuations, i.e.,
$\uu=\UU-\meanUU$ and $\bb=\BB-\meanBB$.
No summation over the index $i$ is assumed.
In \Fig{peffmin} we show the dependence of the minimum value of
$\Peff$ on the Coriolis number (see \Tab{runs2}) and for
different values of the imposed field (see \Tab{runs3}).
It turns out to be relatively insensitive to the value of $\Co$, but it
drops dramatically with increasing strength of the imposed field.

Given that the values of $\Peff$ shown in \Fig{peffmin} for different
values of $\Co$ are almost the same, we conclude that those coefficients
do not depend on $\Omega$.
We can therefore obtain a new set of parameters that is representative
of our model with coronal layer and for all values of $\Omega$, which
is different from our standard representation without a coronal envelope.
This is shown in \Fig{ppeff}.
The values of the parameters $\qpz$ and $\betap$ fit very well with the
data of \cite{WLBKR2015}, shown as triangles.

The new data can be fitted to \Eq{param_beta} by
determining the position and value of the minimum,
$\betamin$ and $\Pmin$, respectively.
Looking at \Fig{ppeff}, we find $\betamin\approx0.3$ and
$\Pmin\approx-0.12$.
We then obtain the fit parameters as \citep{Kemel12a}
\begin{equation}
\betap=\betamin^2\left/\sqrt{-2\Pmin},\right.\;\;
\betastar=\betap+\sqrt{-2\Pmin}.
\end{equation}
This results in the following new set of parameters:
$\qpz=13$ and $\betap=0.18$, which corresponds to $\beta_\ast=0.65$.
We therefore also present in the next section
MFS results based on our model
with this parameter combination.

\begin{figure}[t!]\begin{center}
\includegraphics[width=\columnwidth]{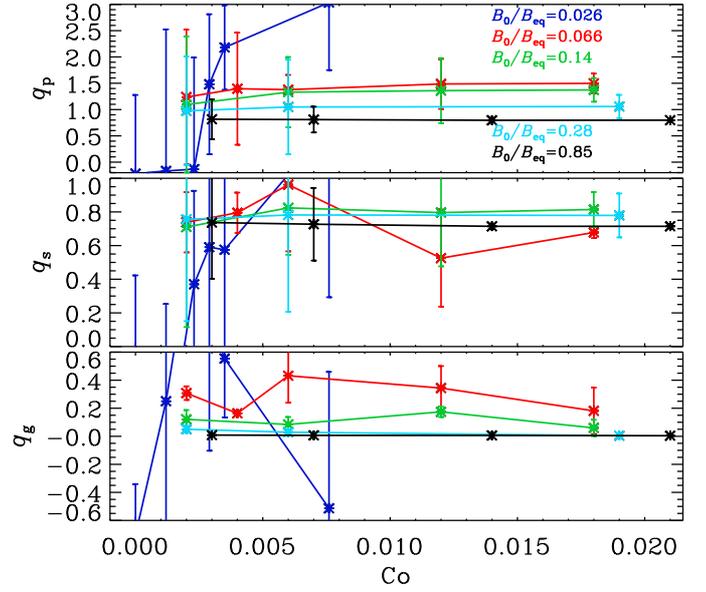}
\end{center}\caption[]{
Dependence $q_{\rm p}$, $q_{\rm s}$, and $q_{\rm g}$ on the Coriolis number for different values of the
imposed field strength, $\beta_0=B_0/\Beqz$ (the curves have been labeled
by an average value), for the runs listed in \Tabs{runs2}{runs3}.
Error bars are estimated using the maximum difference of the total mean
with the means of each third of the time series.
}\label{Qp_co}
\end{figure}

\begin{figure}[t!]\begin{center}
\includegraphics[width=\columnwidth]{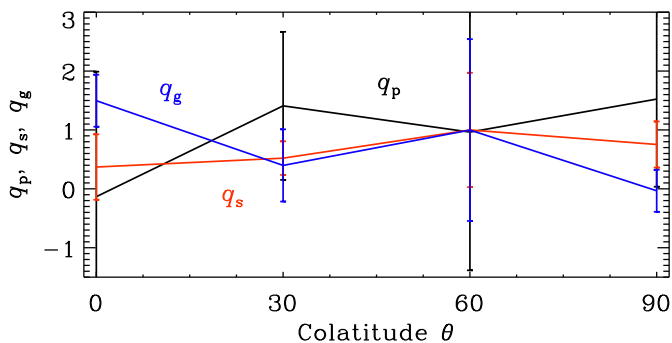}
\end{center}\caption[]{
Dependence $q_{\rm p}$, $q_{\rm s}$ and $q_{\rm g}$ on the colatitude
$\theta$ for Runs~C1--C3 and C5 with Co=0.0023 and $B_0/\Beqz=0.026$.
Error bars are estimated using the maximum difference of the total mean
with the means of each third of the time series.
}\label{Qp_theta}
\end{figure}

Along with $\qp$ and $\Peff$, we also determine $\qs$ and $\qg$ in
\begin{equation}
\Delta\overline\Pi_{ij}=-\qp \delta_{ij} {\mean{\BB}^2\over 2}
+\qs \mean{B_i}\mean{B_j} -\qg {g_i g_j\over g^2} \mean{\BB}^2,
\label{pi-mean}
\end{equation}
resulting in
\begin{equation}
\qs={\Delta\Pi_{xx}-\Delta\Pi_{yy}\over\mean{B_x}^2 -\mean{B_y}^2} ,
\label{q_s}
\end{equation}
\begin{equation}
\qg={1\over\mean{\BB}^2} \,
\left(-\Delta\Pi_{zz}-\qp{\mean{\BB}^2\over 2}
  +\qs\mean{B_z}^2\right),
\label{q_g}
\end{equation}
where ${g_i}$ are components of $\gggg$ which, in our setup, has only a
component in the negative $z$ direction.
In \Fig{Qp_co} we plot the dependencies of $\qp$, $\qs$, and $\qg$
on angular velocity and imposed magnetic field strength.
The values for no and weak rotation are consistent with those
obtained by \cite{WLBKR2015} for a non-roting setup and the same
magnetic field strengths.
Because of the large errors resulting from the strong variation in
space and time, we cannot determine whether the sign is negative
or positive.
For larger imposed magnetic field strengths, the situation becomes more
clear.
There, the errors are significantly larger and $\qp$, $\qs$, and $\qg$
do not strongly depend on rotation.
However, we see a dependence on the imposed magnetic field strength.
In particular, $\qp$ is positive for $B_0/\Beqz\ge0.066$ and
decreases from around $1.5$ to $1.0$ for increasing imposed
magnetic field strength.
Next, $\qs$ is also positive for imposed magnetic fields $B_0/\Beqz\ge0.066$,
but has an inconclusive dependence on the imposed magnetic field.
Finally, $\qg$ is also positive for $B_0/\Beqz\ge0.066$,
it decreases with increasing imposed field until it is nearly zero for
$B_0/\Beqz=0.85$.
For $B_0/\Beqz=0.066$, $\qg$ tends to decrease with increasing angular velocity, but not
for the other magnetic field strengths.
We should keep in mind that the $\qp$ and $\qg$ in \Eq{efforce} are
multiplied by the horizontal averaged magnetic field, which will be
larger for larger imposed magnetic fields.
We also look at the latitudinal dependencies of $\qp$, $\qs$, and $\qg$;
see \Fig{Qp_theta}. We do not find any latitudinal dependence of
these coefficients, mostly because the errors are so large.
In conclusion, we cannot explain the rotational dependence
found in the DNS with just the rotational dependence of $\qp$, $\qs$,
and $\qg$.

\section{MFS results}

\subsection{General aspects}

We now present mean-field calculations with a coronal envelope.
In all cases, we use $\nuT=\etaT=10^{-3}\cs H_\rho$ and $\urms=0.1\cs$,
corresponding to $\kf H_\rho=33$, which is similar to the value
of $30$ used in the DNS.
We first consider the same domain size as in the DNS, i.e., $(2\pi)^2\times3\pi$.
To alleviate finite domain size effects, we also consider a wider and
deeper domain, but with a smaller coronal part that is reduced by half.
We thus also consider $L_x=L_y=6\pi$, $L_z=4\pi$, and $z_{\rm top}=\pi$.

The MFS lack small-scale turbulent motions, so fewer mesh points
can be used.
However, to resolve the vertical density contrast of around 12,000, we
used 288 mesh points in the $z$ direction.
For our domains, are used $192^2\times288$ meshpoints, but we found no
differences in the results when using $96^2\times288$ meshpoints.
For the larger domains, we used $384^3$ meshpoints, but again, with
fewer meshpoints the results would have been sufficiently accurate.
In all cases, we use $B_0/\Beqz=0.1$, because the DNS discussed in
\Sec{DependenceFieldStrength}
showed that for approximately this value, the range in rotation rates over
which NEMPI is still excited would be maximized.

Although we have found that the parameters $\qpz\approx13$ and
$\betap\approx0.65$ describe the DNS best, there are reasons to consider
also other choices.
First, there is no good reason why the parameters $\qpz$ and $\betap$
are so different in different circumstances.
One would have expected them to reflect properties of the turbulence
which is similar in all the different cases.
Second, the response for vertical and horizontal magnetic fields turns
out to be different; see Fig.~7 of \cite{LBKR2014}.
There are also other differences between MFS and DNS that we address
below.
Thus, we cannot expect the two approaches to agree. One
objective is therefore to find out just how well the MFS perform relative
to the DNS.

\begin{figure}[t!]\begin{center}
\includegraphics[width=\columnwidth]{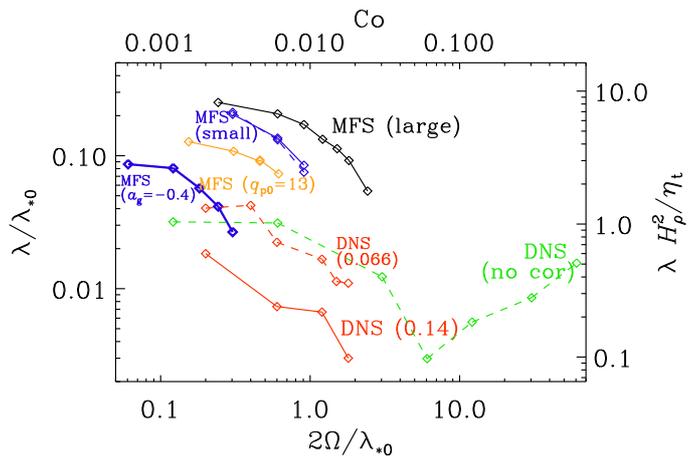}
\end{center}\caption[]{
Dependence of $\lambda/\lambda_{\ast0}$ on $2\Omega/\lambda_{\ast0}$
for $B_0/\Beqz=0.1$ and $\theta=0$ for the MFS using the large (black
line) and small domains (blue solid line), as well as the DNS for
neighboring magnetic field strength of $B_0/\Beqz=0.066$ and 0.14
(dashed and solid red lines, respectively).
The fat blue line denotes the case of a small domain
using $a_{\rm g}=-0.4$ instead of $0$.
The green dashed line denotes the DNS of \cite{LBKR2013a} without coronal
layer and $B_0/\Beqz=0.05$.
The orange line refers to MFS in a small domain with $\qpz=13$ and
$\betap=0.18$, so $\beta_\ast=0.65$.
The blue dashed line adjacent to the blue solid line denotes the results
obtained when neglecting the derivative term of the profile function
$\Theta_w(z)$.
}\label{presults_Omdep}
\end{figure}

\subsection{Growth rates}

In the DNS, the value of $\lambda/\lambda_{\ast0}$ is found to drop by about a
factor of five as $2\Omega/\lambda_{\ast0}$ increases from 0.2 to 3;
see \Fig{presults_Omdep}.
We compare the MFS for $B_0/\Beqz=0.1$ with the DNS at $B_0/\Beqz=0.066$
and $0.14$; see \Tab{runs2}.
We also compare with the DNS of \cite{LBKR2013a} without coronal envelope
using $\beta_\ast=0.33$ and, as in \cite{LBKR2013a}, the value
$\beta_\ast=0.75$, which was suitable for one of their sets of MFS.
This now explains the 2.3 times larger ratio
of $\lambda/\lambda_{\ast0}$ compared to their Fig.~2.
In addition, however, the growth rates of their DNS were incorrectly
scaled by a factor $(\urms/\cs)(\kf/k_1)\approx0.1\times30=3$, which is
now corrected; see the green line of \Fig{presults_Omdep}.
Thus, even in the absence of a corona, the growth rates were
by a factor of about seven larger in the MFS than in the DNS; see
\App{Comparison_Los_etal13} with a corrected version of Fig.~2 of
\cite{LBKR2013a}.

Again, the growth rates of the large-scale instability
in the MFS are significantly larger than those in the DNS.
We speculate that this could be caused by a partial cancelation
(i.e., a decrease of the effective magnetic pressure gradient)
from the $q_{\rm g}$ term in \Eq{efforce}.
To illustrate this possibility, we have overplotted in \Fig{presults_Omdep}
a case with $a_{\rm g}=-0.4$; see \Eq{adef}.
Given that in \Eq{efforce}, only the $\qp$ term comes with a 1/2 factor,
the effective magnetic pressure gradient in the $z$ direction is reduced.
Thus, $\qp/2$ is replaced by $\qp/2+\qg=(1/2+a_{\rm g})\,\qp$, so $\qp$ is
scaled by a factor $1+2a_{\rm g}=0.2$ with respect to the $z$ direction.
We see that the growth rate is now suppressed, but also the maximum
rotation rate for which NEMPI can operate is reduced.
Obviously, a more accurate modeling requires a more detailed knowledge
of the actual form of $q_{\rm g}$, which is likely to be different
from that of $q_{\rm p}$.
We should also point out that, when
calculating $\partial_z q_{\rm g}(\beta)$ in \Eq{efforce},
one of the resulting terms involves the gradient of $\Theta_w(z)$
in \Eq{parameterization}. We have verified that neglecting this term
causes only a very minor change in
the resulting growth rates; cf.\ the solid and dashed blue lines in
\Fig{presults_Omdep}.

Let us discuss the non-monotonic behavior of the
growth rate of the magnetic field as a function of the Coriolis
number; see \Fig{presults_Omdep}.
In corresponding DNS without coronal layer \citep{Jabbari2014},
the increase in the growth rate
at faster rotation rates ($\Co>0.1$ or $2\Omega/\lambda_{\ast0}>10$)
has been explained as a result of large-scale dynamo action;
see Fig.~8 of \cite{Jabbari2014}.
In particular, at larger rotation rates, kinetic helicity is produced by a
combined effect of uniform rotation and density stratified turbulence.
It results in the excitation
of an $\alpha \Omega$ or $\alpha^2\Omega$ dynamo instabilities
and the generation of a large-scale magnetic field.
This causes an increase of the growth rate at larger Coriolis numbers,
which is also observed in the DNS of
\cite{LBKR2013a} and \cite{Jabbari2014}.
This implies that two different instabilities are excited
in the system, i.e., NEMPI at low Coriolis numbers
and the mean-field dynamo instability at larger values of the
Coriolis numbers. This causes a non-monotonic behavior of the
growth rate of magnetic field as the function of the Coriolis
number, observed in \Fig{presults_Omdep}.

A non-trivial evolution of the magnetic field
in rotating turbulence with a coronal envelope
is caused for the following reasons.
In an earlier study by \cite{LBKMR2012}
with rotation, no coronal envelope, and
an imposed horizontal magnetic field,
an analytical expression for the growth rate of NEMPI
has been derived in the framework
of a mean-field approach.
During the magnetic field evolution in the presence of a coronal
envelope, as in the simulation of \cite{WLBKR2013,WLBKR2015} and the
present ones, there is a change in the direction of
the large-scale magnetic field from horizontal at $t=0$ to
nearly vertical after about one turbulent diffusive time.
Therefore, in turbulence with a coronal envelope
one can expect to find a mixture of
effects caused by the horizontal and vertical magnetic fields.

The growth rates and saturation mechanisms of NEMPI for horizontal and
vertical fields are very different \citep[see review by][]{BRK2016}.
For horizontal fields, NEMPI saturates rapidly
in the nonlinear stage of the magnetic field evolution
due to the ``potato-sack" effect.
This means that a local increase of the magnetic field
causes a decrease of the negative effective magnetic pressure,
which is compensated for by enhanced gas pressure.
This leads to an enhanced gas density, so the gas
is heavier than its surroundings and sinks.
This effect removes horizontal magnetic flux structures from regions
in which NEMPI is excited.
For a vertical magnetic field, the heavier fluid moves
downward along the field without affecting the flux tube, so
that NEMPI is not stabilized by the potato sack effect.
In this case the operation of NEMPI results
in the formation of strong concentrations \citep[see][]{BKR2013,BRK2016}.

In the non-rotating cases,
the growth rates of the large-scale magnetic field measured in DNS
with coronal envelope and horizontal initial field \citep{WLBKR2015},
and with vertical magnetic field but without corona \citep{BKR2013}, are the same.
This is an indication that the growth rates of the large-scale magnetic field
in the simulations with coronal envelope are determined by the evolution
of the vertical field.
On the other hand,
the fact that the BRs dissolve after a few turbulent diffusion
times is more similar to the behavior of NEMPI with a horizontal imposed
magnetic field.
Therefore, the evolution of the magnetic structures
in the system with
coronal envelope is non-trivial and cannot easily be described with our
current mean-field models.
Therefore, also explaining the rotational dependency of growth rates
in this setup suffers from the mixture of effects.

\begin{figure}[t!]\begin{center}
\includegraphics[width=\columnwidth]{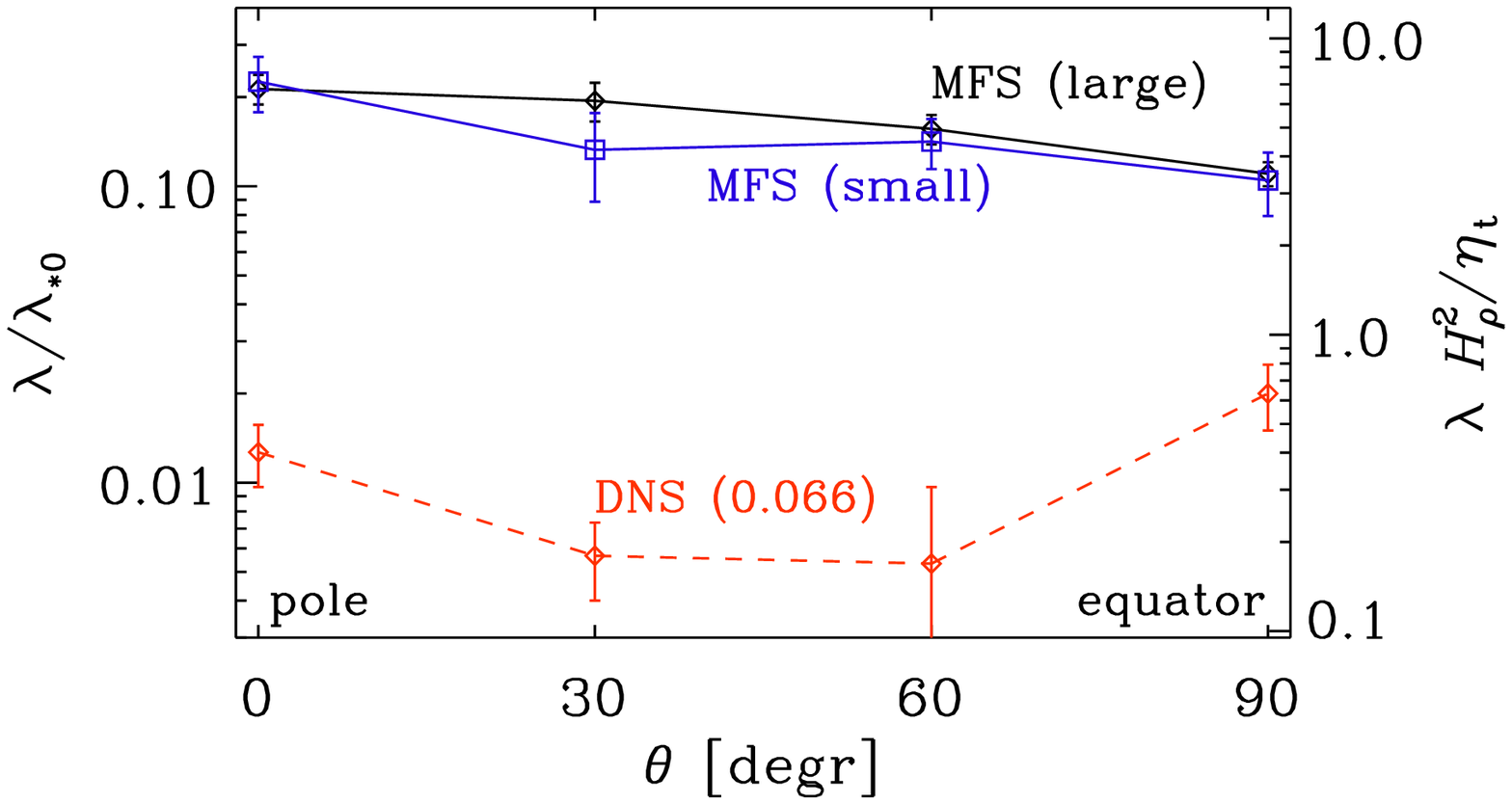}
\end{center}\caption[]{
Dependence of $\lambda/\lambda_{\ast0}$ on $\theta$ for
$B_0/\Beqz=0.1$ and $\Co=0.006$.
}\label{plam_theta}
\end{figure}

\subsection{Latitudinal dependence}

At slower rotation, a decrease in $\lambda/\lambda_{\ast0}$ can be seen
as $\theta$ increases from $\theta=0$ at the poles to $\theta=90\degr$
at the equator; see \Fig{plam_theta}.
In both plots we also show the growth rates for the larger
domain, which are found to be enhanced by a factor of two when
$2\Omega/\lambda_{\ast0}\approx1$.
These values are about an order of magnitude larger than those
for the DNS, but have otherwise a similar functional dependence
on $\Omega$ and $\theta$.
The reason for the difference between DNS and MFS is not entirely clear.
It is possible that the mean-field parameter $\beta_\ast$ is smaller
than what was previously found for simulations with coronal envelope.
There is also the possibility that $\beta_\ast$ decreases with increasing
angular velocity, which is what R\"udiger (private communication) found,
although our present simulations presented in \Fig{peffmin} and earlier
ones of \cite{LBKR2013a} did not give such indications.

\subsection{Inclined surface structures}
\label{InclinedSurfaceStructures}

\begin{figure}[t!]\begin{center}
\includegraphics[width=\columnwidth]{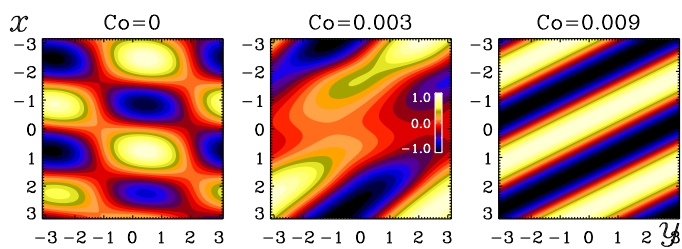}
\end{center}\caption[]{
Slices of $\meanB_z(x,y,0,t_\ast)$ through the surface $z=0$
at times $t_\ast$ during the linear growth phase of NEMPI
for $B_0/\Beqz=0.1$ and three values of $\Co$ using the smaller
domain size of $(2\pi)^2\times3\pi$.
In each panel, the magnetic field is scaled to the maximum value.
}\label{pslice_comp2}
\end{figure}

\begin{figure}[t!]\begin{center}
\includegraphics[width=\columnwidth]{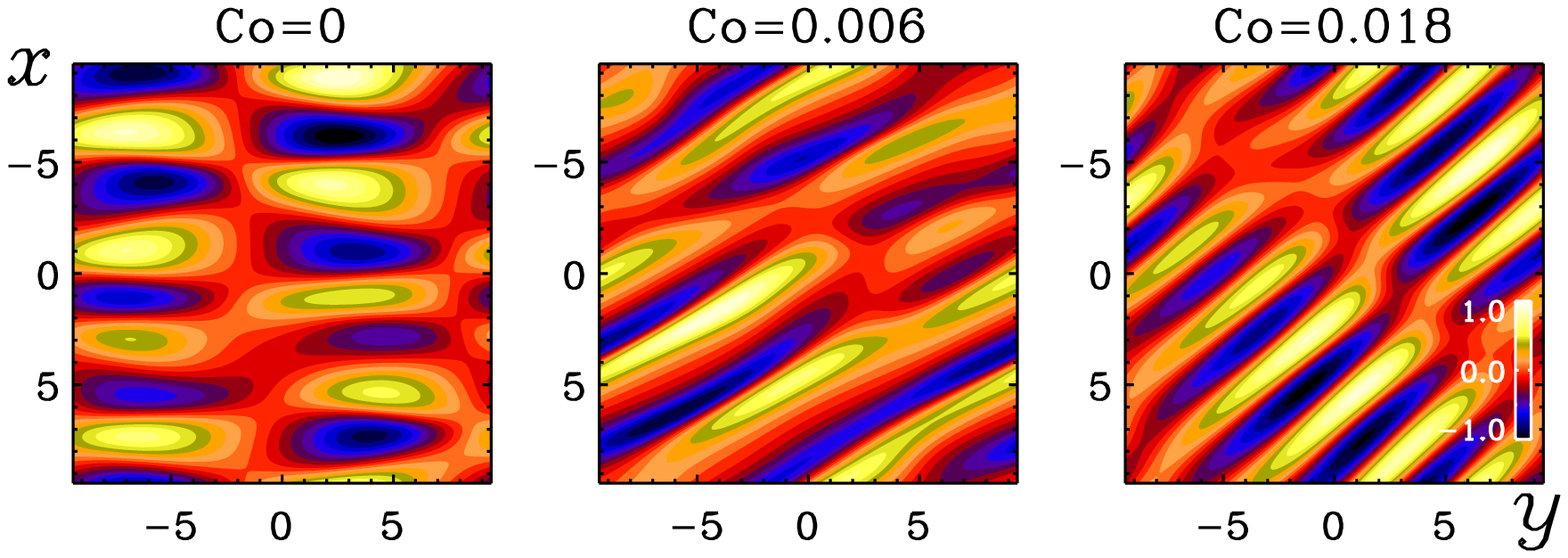}
\end{center}\caption[]{
Similar to \Fig{pslice_comp2}, but for the large domain,
$(6\pi)^2\times4\pi$.
}\label{pslice_comp}
\end{figure}

Next, we show slices of $\meanB_z(x,y,0,t_\ast)$ along the surface $z=0$
at a chosen time $t_\ast$ during the linear growth phase of NEMPI for three
values of $\Co$ using domain sizes of $(2\pi)^2\times3\pi$
(\Fig{pslice_comp2}) and $(6\pi)^2\times4\pi$ (\Fig{pslice_comp}).
In the linear phase, when the magnetic field fluctuations are still
growing exponentially in time, only relative values are of physical
interest.
We therefore present in the following images where the magnetic field
is normalized by the maximum value.
As in the DNS, the imposed magnetic field points in the $y$ direction.
It turns out that rotation not only tends to make the structures inclined
relative to the direction of the imposed magnetic field,
but it also leads to higher wavenumbers of the structures.
\FFig{pslice_comp} shows that the number of nodes in the $x$ direction,
which is perpendicular to the magnetic field, remains about constant
($k_x L_x/2\pi=4$), while that in the $y$ direction along the magnetic
field increases from $k_y L_y/2\pi=1$ (for $\Co=0$) to 2 (for $\Co=0.006$)
and 4 (for $\Co=0.018$).

\begin{figure}[t!]\begin{center}
\includegraphics[width=\columnwidth]{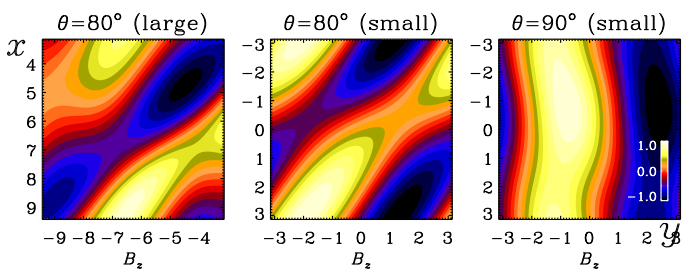}
\end{center}\caption[]{
Similar to \Fig{pslice_comp2}, but for two runs with
$\theta=80\degr$ for a 1/3 section of the large domain
and the full section of the smaller domain,
as well as a run with $\theta=90\degr$.
}\label{pslice_comp80}
\end{figure}

In both the DNS and the MFS, the orientation of the inclination is the same
and it is opposite to what is seen in Joy's law.
The runs presented here apply of course only to the poles, but even
at lower latitudes (e.g., at $30\degr$ latitude, corresponding to
$\theta=60\degr$) do we find the same anti-Joy's law orientation of
the tilt.
This is shown in \Fig{pslice_comp80}, where we compare two runs with
$\theta=80\degr$ (close to the equator) for a section of the large
domain and the full section of the smaller domain, as well as a run
with $\theta=90\degr$ (at the equator).
At the equator, the inclination angle with respect to the toroidal
direction is $90\degr$, which agrees with what was found in
\cite{LBKMR2012}.
However, slightly away from the equator, at $\theta=80\degr$, the
inclination angle is already $45\degr$.
This is not an artifact of having chosen a small domain, because even
for a three times larger domain, the same inclination angle is found.
We thus remain puzzled about this finding and hope to be able to
return to it as further numerical and analytic results can be obtained
and assessed.

One might speculate that the reason for the difference to Joy's law has
to do with the expansion of rising structures whereas NEMPI structures
are caused by contraction, which leads to the opposite tilt.
Thinking again of possible applications to the Sun, one may therefore
wonder whether the effect of flux concentrations in NEMPI, which must also
be responsible for causing the tilt, operate on relatively small scales
and might be responsible for causing sunspot rotation
\citep{Eve09,Kem10,Pev12,Stu15}, for example.
In a distributed solar dynamo scenario, the tilt of active regions is
primarily caused either by differential rotation \citep{Bra05} or simply
by the sign of the mean latitudinal field $\meanB_\theta$ relative to
that of the azimuthal field $\meanB_\phi$ \citep{Jabbari15}.

\subsection{Emergence of solitary structures}

\begin{figure}[t!]\begin{center}
\includegraphics[width=\columnwidth]{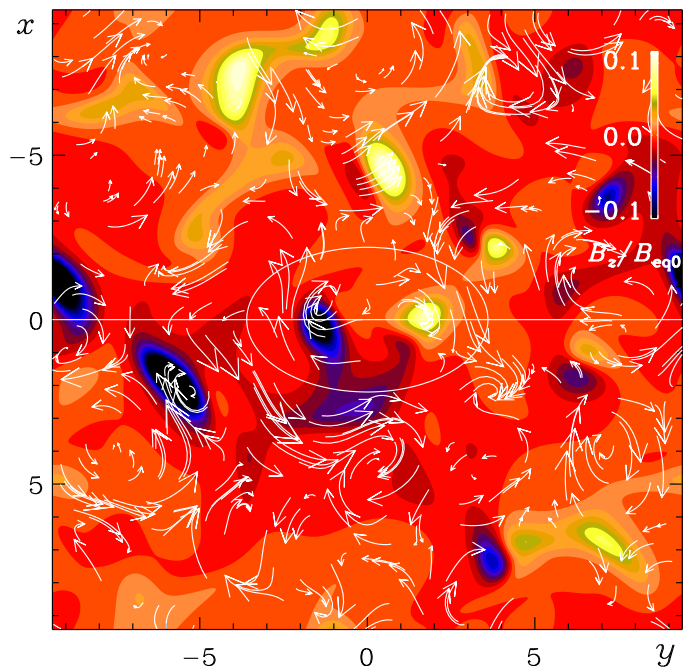}
\end{center}\caption[]{
A slice of $\meanB_z(x,y,z_\ast,t)$ (color coded) together with
velocity field vectors (white streak lines) through $z_\ast=0$ at
$t_\ast/\tautd=40$ during the saturated state for the large domain
using $\Co=0.006$.
The white horizontal line through $x=0$ marks the position
of a BR near $y=0$, which is discussed separately.
The ellipse marks the position of the BR discussed in the text.
}\label{pxy_slices_1h}
\end{figure}

The magnetic field patterns in \Figss{pslice_comp2}{pslice_comp80}
are more or less regular.
This is basically because those pictures were taken from the linear
phase of the run.
In \Fig{pxy_slices_1h} we show a visualization of $\meanB_z$ along with
horizontal flow vectors $(\meanU_x,\meanU_y)$ through the surface for
an arbitrarily chosen time $t_\ast/\tautd=40$ during the saturated state
for the large domain using $\Co=0.006$.
We can now clearly see solitary structures in the form of isolated spots.
However, with a few exceptions, most of these structures lack the distinct
bipolarity seen in the DNS.
Note also that the mean flow is mostly circular around each spot rather
than a convergent inflow, as seen in the DNS.

In \Fig{pxy_slices_1h} we mark with a white horizontal line in
the toroidal direction through $x=0$ the position of a BR near $y=0$.
Unlike the DNS, the separation of the two polarities is rather large
(about $3\,H_\rho$) and there are other spots in almost the same
distance.
Thus, it is not clear that these two polarities are connected to each
other.
There is also no clear indication of BRs.
To examine this further, we present in the next section a side view of
this BR.

\subsection{Side view of BRs}

\begin{figure}[t!]\begin{center}
\includegraphics[width=\columnwidth]{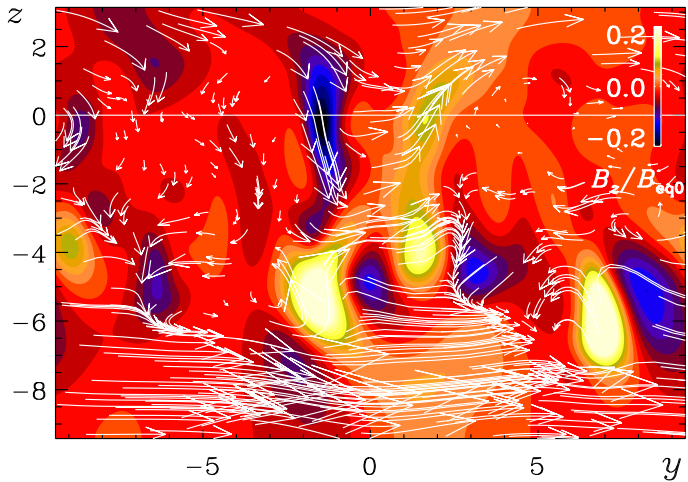}
\end{center}\caption[]{
A slice of $\meanB_z(x_\ast,y,z,t)$ (color coded) together with
magnetic field vectors (white streak lines) through $x_\ast=0$
at an arbitrarily chosen time during the saturated state for
the large domain.
The surface at $z=0$ is shown as a white horizontal line.
}\label{pyz_slices_1h}
\end{figure}

In \Fig{pyz_slices_1h} we show a longitudinal cross-section of
the magnetic field through $x_\ast=0$ at the same time as
\Fig{pxy_slices_1h} during the saturated state.
Magnetic flux concentrations are seen to occur at a depth of
$z/H_\rho\approx-6$, which is well below the surface.
Near the surface, on the other hand, there is only a relatively
small number of vertical magnetic flux structures
that seem to close upon themselves over relatively large horizontal distances.
We recognize the positive and negative polarities at
$y/H_\rho\approx\pm1.5$ and the negative one at $y/H_\rho\approx-8$
in both \Figs{pxy_slices_1h}{pyz_slices_1h}.
Conversely, over short distances, bipolar magnetic flux
structures separate above the surface, which is consistent
with them being the result of a localized subduction of
a horizontal flux structure.

\begin{figure}[t!]\begin{center}
\includegraphics[width=\columnwidth]{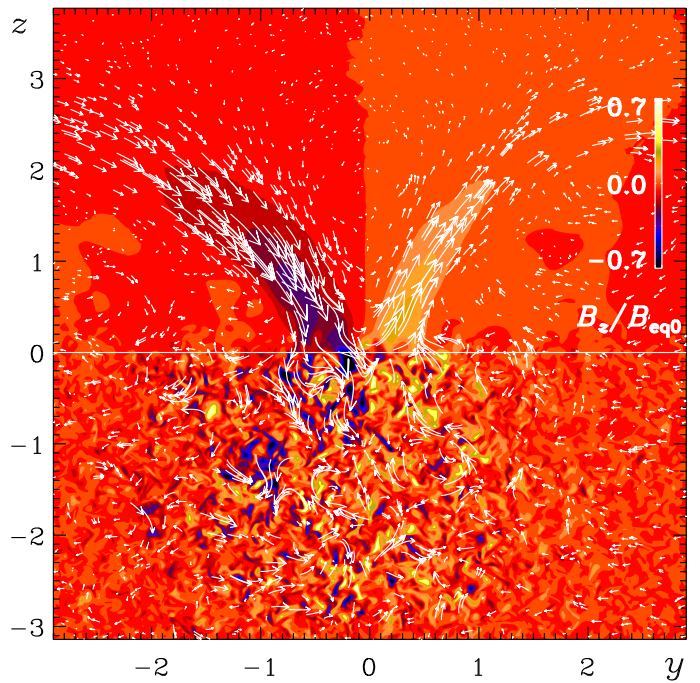}
\end{center}\caption[]{
Comparison with DNS showing $B_z(x_\ast,y,z,t)$ (color coded) together
with vectors of Fourier-filtered magnetic field vectors ($k<\kf/2$)
superimposed for Run~A4 at the time $t/\tautd=0.91$ through $x_\ast=0.5$.
}\label{pyz_slice}\end{figure}

To compare with DNS results, we show in \Fig{pyz_slice} a similar plot
of $B_z$ together with Fourier-filtered magnetic field vectors in the
same plane.
A major difference to \Fig{pyz_slices_1h} is the absence of
significant horizontal field in the deeper parts.
However, since this horizontal field in the MFS is so deep down
($z/H_\rho\la-8$), it is unclear whether it plays any role in
explaining the difference in, for example, the growth rates
between DNS and MFS seen in \Fig{presults_Omdep}.
On the other hand, in the deeper parts of the MFS, there are magnetic
structures of significant strength, which are not so prominent in the DNS.
This is an important difference that would affect global comparisons of,
for example, the growth rates of structures shown in \Fig{presults_Omdep}.

\section{Discussion}

Our calculations have been performed using idealizing circumstances
such as forced turbulence and an isothermal equation of state.
This is in many ways different from turbulence in the Sun,
which is driven by convection.
Nevertheless, some tentative conclusions can be drawn regarding
possible applications to sunspot formation.
In the surface layers of the Sun, the turnover time $\tauto=\HP/\urms$
based on the pressure scale height $\HP$ is around $5\min$.
Assuming $k_1\HP=1$, the turbulent diffusive time scale is
related to this via
\EQ
\tautd=\HP^2/\etatz=3\kf\HP^2/\urms=3\kf\HP\tauto.
\EN
In the simulations, we have $\kf\HP\approx\kf/k_1=30$,
so $\tautd$ would be about 90 times longer than $\tauto$,
i.e., about 8 hours.
This would appear suitable in view of applications to the Sun as well,
where the formation time of sunspots is of a similar order.

Although the presence of our simplified corona has the effect of allowing
BRs to form that reach a significant fraction of the equipartition value
with respect to the turbulence, these structures can no longer form when
the Coriolis number exceeds a critical value of about 0.02.
This value is rather small.
However, given that the growth rate of NEMPI does not scale
with the inverse turnover time $\tauto^{-1}$, but with the
turbulent-diffusive
time $\tautd^{-1}$, a meaningful measure of the rotation rate in this
context could also be the square root of the {\it turbulent} Taylor number,
$\Tat^{1/2}=2\Omega/\nut k_1^2=3\,(\kf/k_1)^2\Co$,
where we took into account that the turbulent viscosity $\nut$ is equal
to the turbulent magnetic diffusivity $\etat$ \citep{YBR03,KR94}
and given by $\etat\approx\urms/3\kf$ \citep{SBS08}.
The values of $\Ta^{1/2}$ are typically of the order of 10 when NEMPI
begins to be suppressed.
Using our estimate of $\tautd=8\,{\rm hours}$ for the solar surface,
we find $\Ta^{1/2}=0.2$, which is well below our critical value of 10.

In the standard mixing length theory of convection, the value of $\kf\HP$
is estimated to be around 6--7 \citep{Kemel13a}, but it is about $30$ in
the present simulations.
Earlier work showed that NEMPI would not work for $\kf\HP$ much below 15
\citep{BKKR12}.
On the other hand, the actual value in the Sun is unclear given that
the findings of \cite{HDS12} did not confirm turbulent velocities
in the Sun at the expected levels.
A possible resolution to this problem might be the idea that the relevant
scales of the energy-carrying eddies is much smaller than the inverse
pressure scale height \citep{Bra15}.
This would also help making NEMPI more powerful.
However, this issue is controversial in view of results by \cite{Greer15},
which showed that the turbulent flows in the Sun might actually be
just as large as assumed in standard mixing length theory.
Thus, the possibility of NEMPI being responsible for the production
of sunspots is being favored particularly in the scenario envisaged
by \cite{HDS12}.

An additional complication is that at the solar surface, radiation
plays an important role.
Mean-field models with radiation transport \citep{PB18} have shown
that the relevant length scale of NEMPI can drop significantly
below the value found for isothermal and isentropic stratifications.
It is possible, however, that this result is a consequence of not
having included the convective flux in such a model.
In the Sun, virtually 100\% of the energy is transported by convection
almost immediately beneath the surface, so radiation should be
completely unimportant below the surface.
In addition, ionization dynamics can strongly exaggerate the effects
of cooling near the surface.

\section{Conclusions}

Our work has confirmed that NEMPI cannot be exited at Coriolis numbers
above a critical value that can be as low as $0.02$ or so.
The presence of an upper coronal layer was previously found to make the
appearance of structures more prominent.
However, rotation seems to affect the growth
rates more strongly with than without a coronal envelope.
In the bulk of the solar convection zone, the Coriolis number is of the
order of unity and above, but this is not the case in the surface layers,
where the convective time scale is much shorter than the solar rotation
period of 25 days.
So this may not really be a problem for applications of NEMPI to sunspot
formation.

A more severe problem for astrophysical applications of NEMPI are the
moderate magnetic field strengths that can presently be achieved with
NEMPI.
This suggests that some essential physics is still missing.
An important ingredient of sunspots physics is convection and its
suppression in the presence of magnetic fields.
A number of aspects such as radiation and ionization physics, taken in
isolation, have not yet produced more favorable conditions for NEMPI
\citep{BhatBrandenburg15,PB18}.

It is important to realize that the difficulty in explaining the
spontaneous formation of sunspot-like magnetic flux concentrations is
not really alleviated by invoking the rising flux tube scenario.
The problem here is that any flux tube rising from some depth to the
surface will expand and therefore weaken and so some mechanism for
reamplification is needed.
Invoking therefore a suitable model for convection,
or possibly the inclusion of a magnetic
suppression of turbulent radiative diffusion as
suggested by \cite{KitchatinovMazur2000},
might be an important next step.

\begin{acknowledgements}
J.W.\ acknowledges funding by the Max-Planck/Princeton Center for
Plasma Physics and funding from the People Program (Marie Curie
Actions) of the European Union's Seventh Framework Programmed
(FP7/2007-2013) under REA grant agreement No.\ 623609.
This work has been supported in part by
the NSF Astronomy and Astrophysics Grants Program (grant 1615100),
the Research Council of Norway under the FRINATEK (grant 231444),
the Swedish Research Council (grant 2012-5797),
and the University of Colorado through its support of the
George Ellery Hale visiting faculty appointment.
I.R.\ acknowledges the hospitality of NORDITA
and Max Planck Institute for Solar System Research in G\"ottingen.
We acknowledge the allocation of computing resources
provided by the Swedish National Allocations Committee at the Center for
Parallel Computers at the Royal Institute of Technology in Stockholm.
This work utilized the Janus supercomputer, which is supported by the
National Science Foundation (award number CNS-0821794), the University
of Colorado Boulder, the University of Colorado Denver, and the National
Center for Atmospheric Research. The Janus supercomputer is operated by
the University of Colorado Boulder.
Additional simulations have been carried out on supercomputers at
GWDG, on the Max Planck supercomputer at RZG in Garching, in the
facilities hosted by the CSC---IT Center for Science in Espoo,
Finland, which are financed by the Finnish ministry of education.
\end{acknowledgements}

\bibliography{ref}
\bibliographystyle{aa}

\appendix

\section{Comparison of growth rates in DNS and MFS of \cite{LBKR2013a}}
\label{Comparison_Los_etal13}

\begin{figure}[h!]\begin{center}
\includegraphics[width=\columnwidth]{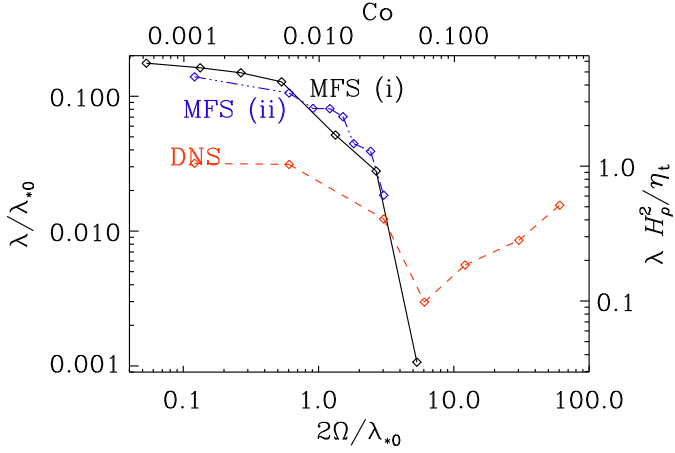}
\end{center}\caption[]{
Corrected version of Fig.~2 of \cite{LBKR2013a} showing the
dependence of $\lambda/\lambda_{\ast0}$ on $2\Omega/\lambda_{\ast0}$
for DNS (red dashed line), compared with MFS~(i) where $\qpz=20$ and
$\betap=0.167$ (black solid line), and MFS~(ii) where $\qpz=32$
and $\betap=0.058$ (blue dash-dotted line).
}\label{presults_Omdep_Los_etal13}
\end{figure}

In \cite{LBKR2013a}, the growth rates were accidently scaled by
a factor $(\urms/\cs)(\kf/k_1)\approx0.1\times30=3$.
In addition, they used $\beta_\ast=0.75$, which was suitable for one of
their sets of MFS, but not for the other.
Therefore, the growth rates of their MFS exceeded those of their DNS by
a factor of about seven.
The corrected version of their Fig.~2 is shown in
\Fig{presults_Omdep_Los_etal13}.

\end{document}